\documentclass[useAMS,usenatbib,twocolumns]{mn2e}
\usepackage{graphicx}
\usepackage[english]{babel}
\usepackage{multirow}
\usepackage{lscape}
\usepackage{longtable}
\usepackage{natbib}
\usepackage[usenames,dvipsnames]{color}
\usepackage{float}
\usepackage{url}
\usepackage[T1]{fontenc}
\usepackage{arydshln}
\usepackage{float}
\usepackage{amsmath}
\usepackage{amssymb}
\usepackage{subfigure}
\usepackage{flafter}
\usepackage{adjustbox}
\usepackage{blindtext}
\usepackage{color}
\usepackage[pdfborder={0.1 0.1 0.7}]{hyperref}
\hypersetup{pdflinkmargin=0.1mm}
\def\LaTeX{L\kern-.36em\raise.3ex\hbox{a}\kern-.15em
    T\kern-.1667em\lower.7ex\hbox{E}\kern-.125emX}

\newcommand{\OII}{$\left[\mathrm{O\textrm{\textsc{ii}}}\right]\,$}

\title[ELG clustering at $z\sim0.8$]{Clustering properties of $g$-selected galaxies at $z\sim0.8$}

\author[Favole et al. 2015]
{\parbox[t]{\textwidth}{\vspace{-0.7cm}Ginevra Favole$^{1,2}$\thanks{E-mail: g.favole@csic.es}, Johan Comparat$^{1,2,5}$\thanks {Severo Ochoa IFT Fellow}, Francisco Prada$^{3,1,2,4}$, Gustavo Yepes$^5$, Eric Jullo$^{6}$, Anna Niemiec$^{6}$, Jean-Paul Kneib$^{7,6}$,  Sergio A. Rodr\'iguez-Torres$^{1,2,5}$\thanks {Campus de Excelencia Internacional UAM/CSIC Scholar}, Anatoly Klypin$^{8}$, Ramin A. Skibba$^{9}$, Cameron K. McBride$^{10}$, Daniel J. Eisenstein$^{10}$, David J. Schlegel$^{3}$, Sebasti\'an E. Nuza$^{11}$, Chia-Hsun Chuang$^{1,2}$\thanks{MultiDark Fellow}, Timoth\'ee Delubac$^{7}$, Christophe Y\`eche$^{12}$, Donald P. Schneider$^{13,14}$} 
\vspace*{15pt}\\ 
$^1$Instituto de F\'{i}sica Te\'{o}rica (IFT) UAM/CSIC, Universidad Aut\'{o}noma de Madrid, Cantoblanco, E-28049 Madrid, Spain\\
$^2$Campus of International Excellence UAM/CSIC, Cantoblanco, E-28049 Madrid, Spain\\
$^3$Lawrence Berkeley National Laboratory, 1 Cyclotron Road, Berkeley, CA, 94720, USA\\
$^4$ Instituto de Astrof\'isica de Andaluc\'ia (CSIC), Granada, E-18008, Spain \\
$^5$Departamento de F\'isica Te\'orica M-8, Universidad Aut\'onoma de Madrid, Cantoblanco, 28049 Madrid, Spain\\
$^6$Laboratoire d'Astrophysique de Marseille - LAM, Universit\'e d'Aix-Marseille \& CNRS, UMR7326, F-13388 Marseille, France\\ 
$^7$Laboratoire d'Astrophysique, Ecole Polytechnique F\'{e}d\'{e}rale de Lausanne (EPFL), Observatoire de Sauverny, CH-1290 Versoix, Switzerland \label{EPFL}\\
$^8$Astronomy Department, New Mexico State University, MSC 4500, PO Box 30001, Las Cruces, NM 880003-8001, USA\\
$^9$Center for Astrophysics and Space Sciences, University of California, 9500 Gilman Drive, San Diego, CA 92093, USA\\
$^{10}$Harvard-Smithsonian Center for Astrophysics, 60 Garden Street, Cambridge, MA 02138, USA\\
$^{11}$Leibniz-Institut f\"{u}r Astrophysik Potsdam (AIP), An der Sternwarte 16, 14482, Potsdam, Germany\\
$^{12}$CEA, Centre de Saclay, IRFU, 91191, Gif-sur-Yvette, France\\
$^{13}$Department of Astronomy and Astrophysics, The Pennsylvania State University, University Park, PA 16802\\
$^{14}$Institute for Gravitation and the Cosmos, The Pennsylvania State University, University Park, PA 16802\\
\vspace{-0.9cm}
}
\date{  }
\begin{document}
\pagerange{\pageref{firstpage}--\pageref{lastpage}} \pubyear{2015}
\maketitle

\begin{abstract}

\noindent Current and future large redshift surveys, as the Sloan Digital Sky Survey IV extended Baryon Oscillation Spectroscopic Survey (SDSS-IV/eBOSS) or the Dark Energy Spectroscopic Instrument (DESI), will use Emission-Line Galaxies (ELG) to probe cosmological models by mapping the large-scale structure of the Universe in the redshift range $0.6 < z < 1.7$. 
With current data, we explore the halo-galaxy connection by measuring three clustering properties of $g$-selected ELGs as matter tracers in the redshift range $0.6 < z < 1$:
(i) the redshift-space two-point correlation function using spectroscopic redshifts from the BOSS ELG sample and VIPERS; (ii) the angular two-point correlation function on the footprint of the CFHT-LS; (iii) the galaxy-galaxy lensing signal around the ELGs using the CFHTLenS.
We interpret these observations by mapping them onto the latest high-resolution MultiDark Planck N-body simulation, using a novel (Sub)Halo-Abundance Matching technique that accounts for the ELG incompleteness. ELGs at $z\sim0.8$ live in halos of $(1\pm 0.5)\times10^{12}\,h^{-1}$M$_{\odot}$ and 22.5$\pm2.5$\% of them are satellites belonging to a larger halo. The halo occupation distribution of ELGs indicates that we are sampling the galaxies in which stars form in the most efficient way, according to their stellar-to-halo mass ratio.
\end{abstract}
\begin{keywords}
 galaxies: distances and redshifts \textemdash\;galaxies: haloes \textemdash\;galaxies: statistics \textemdash\;cosmology: observations \textemdash\;cosmology: theory \textemdash\;large-scale structure of Universe
\end{keywords}


\section{Introduction}
\label{Sec:Intro}
By investigating the properties of galaxy clustering within the cosmic web, it is possible to constrain cosmology and infer the growth of structure and the expansion history of the Universe \citep{2013PhR...530...87W}. In fact, galaxy clustering measurements using last-generation large-volume redshift surveys, as the Sloan Digital Sky Survey \citep[SDSS;][]{York2000, Gunn2006, Smee2013} and the SDSS-III Baryon Oscillation Spectroscopic Survey \citep[BOSS;][]{Eisenstein2011, Dawson2013} provide robust information about both the evolution of galaxies and the cosmological framework in which these complex structures live. In order to interpret such measurements, we need to understand the relation between the theory-predicted dark matter field and its luminous counterpart i.e., the discrete galaxy map \citep{2002PhR...372....1C}. 

Luminous low-redshift galaxies have already been connected to their dark matter halos in a precise manner, through weak lensing and clustering analysis as a function of galaxy luminosity and stellar mass. 
\cite{2004ApJ...600..681B}, \cite{2011ApJ...736...59Z} and \cite{2015arXiv150507861G} measured the clustering properties of the SDSS ``blue cloud'' and ``red sequence'' in the local Universe (SDSS median redshift $z\sim0.1$; \citeauthor{Abaz2009} \citeyear{Abaz2009}), as a function of magnitude and color. Their results show that at a given luminosity, the blue sample has a lower clustering amplitude and a smaller correlation length compared to the red one.

\cite{2014MNRAS.441.2398G} investigated the clustering luminosity and colour dependence of BOSS CMASS DR10 \citep{2014MNRAS.441...24A}, and found that more luminous galaxies are more clustered and hosted by more massive halos. For luminous red galaxies (LRGs), these masses are $\sim 10^{13}-10^{14}h^{-1}$M$_{\odot}$, at fixed luminosity, progressively redder galaxies are more strongly clustered on small scales, which can be explained by having a larger fraction of these galaxies in the form of satellites in massive haloes. \cite{Favole2015} measured galaxy clustering in the BOSS CMASS DR11 \citep{2014MNRAS.441...24A} sample at $z>0.55$ as a function of color, and proposed a new statistic to extract robust information about small-scale redshift-space distortions and large-scale galaxy bias. Consistent with many previous results \citep[e.g.,][]{Wang2007, Zehavi2005b, Swanson2008}, they found that, compared to the blue population, red galaxies reside in more massive halos, show a higher clustering amplitude, large-scale bias and peculiar velocities. 

This type of clustering analysis has recently been extended to higher redshifts thanks to the VIMOS Public Extragalactic Survey \citep[VIPERS; ][]{2014A&A...566A.108G, 2014A&A...562A..23G} and DEEP2 survey \citep{2013ApJS..208....5N}. Compared to DEEP2, VIPERS has a much larger volume but has a lower redshift limit however, the signal-to-noise ratio in its spectroscopic measurements is higher. Using VIPERS data, \citet{2013A&A...557A..17M} measured the clustering properties of galaxies at redshift $z=0.8$ as a function of their luminosity and stellar mass, and found that the clustering amplitude and the correlation length increase with these two quantities; see also the PRIsm MUlti-object Survey (PRIMUS) results by \cite{2015ApJ...807..152S} and \cite{2015arXiv150201348B}. 
\cite{2013ApJ...767...89M} measured the clustering of the red sequence and the blue cloud at $z=0.9$, as a function of their stellar mass and star formation history, using DEEP2 data. They argued that blue galaxies are more clustered in the local Universe than at $z=0.9$, and red galaxies are much more clustered locally than at high redshift. They also suggested that the clustering trend observed with star formation rate (SFR) can be explained mostly by the correlation between stellar mass and clustering amplitude for blue galaxies. \cite{Coil2008} studied the DEEP2 clustering dependence on color and luminosity, and found that the dependence on color is much stronger than with luminosity, and is as strong with color at $z\sim1$ as locally. They claimed no dependence of the clustering amplitude on color for galaxies in the red sequence, but a significant dependence for galaxies within the blue cloud. 
\cite{2008MNRAS.383.1058C} investigated the connection between star formation (SF) and environment in DEEP2 data at $z\sim0.1$, and $z\sim1$. Their results indicate that, locally, galaxies in regions of higher overdensity have lower star formation rates (SFRs), and their stars form more slowly than in their counterparts in lower density regions. At $z\sim1$, this SFR-overdensity relation is inverted; this is in part due to a population of bright, blue galaxies in dense environments, which lacks a counterpart in the local Universe, and is thought to evolve into members of the red sequence from redshift 1 to 0.

The combination of clustering with weak galaxy-galaxy lensing (see e.g., \citeauthor{1999elss.conf..213B} \citeyear{1999elss.conf..213B}) allows one to gain insight on the large-scale structure formation, and directly probe the stellar-to-halo mass relation \citep[SHMR;][]{2011ApJ...738...45L}. 
The galaxy-halo connection has been measured at $z<1$ by \cite{2012ApJ...744..159L}, \cite{2015arXiv150200313S}, and \cite{2015MNRAS.449.1352C}, using three different weak lensing surveys (COSMOS: \cite{Scoville2007}; CFHT-Stripe82 and CFHTLenS\footnote{http://www.cfht.hawaii.edu/Science/CFHTLS/}: \cite{Heymans2012, Erben2013}); all obtained consistent results. 
\cite{2012ApJ...744..159L} performed the first joint analysis of galaxy-galaxy weak lensing, galaxy clustering, and galaxy number densities using COSMOS data, and provided robust constraints on the shape and redshift evolution of the SHAM relation in the redshift range $0.2<z<1$. At low stellar mass, the halo mass scales proportionally to $M_{\star}^{0.46}$; this scaling does not evolve significantly with redshift. At $M_{\star}>5\times 10^{10}$M$_{\odot}$, the SHMR rises sharply, causing the stellar mass of a central galaxy to become a poor tracer of its parent halo mass. Combining observations in the CFHT-LenS/VIPERS field from the near-UV to the near-IR, \cite{2015MNRAS.449.1352C} found that the SHMR for the central galaxies peaks at $M_{h,peak}=(1.9^{+0.2}_{-0.1}\times 10^{12}$M$_{\odot})$, and its amplitude decreases as the halo mass increases.
\cite{2014MNRAS.444..729H} presented new measurements of the galaxy two-point correlation function and the galaxy-galaxy lensing signal from SDSS, as a function of color and stellar mass, and demonstrated that the age-matching model \citep{2013MNRAS.435.1313H}, which states that older halos tend to host galaxies with older stellar populations, exhibits remarkable agreement with these and other statistics of low-redshift galaxies.

Current (Sub)Halo-Abundance Matching \citep[SHAM;][]{Conroy2006, Trujillo2011, Klypin2013, 2013MNRAS.432..743N} and Halo Occupation Distribution \cite[HOD;][]{Berlind2002, Kravtsov2004, Zheng2005, Zheng2007} models correctly reproduce the clustering measurement mentioned above. SHAM maps observed galaxies onto dark matter halos directly from N-body cosmological simulations, according to a precise monotonic  correspondence between halo and galaxy number densities. The HOD method is an analytical prescription to populate simulated halos with galaxies, in which the assignment is perfomerd by interpolating the halo occupation distribution at the values of the desired halo masses. In this sense, the SHAM approach  returns a model which is built directly on the considered simulation box.

Next generation high-redshift surveys as SDSS-IV/eBOSS (Dawson et al. 2015, in prep.), Subaru Prime Focus Spectrograph \citep[PFS;][]{2012SPIE.8446E..0YS, 2014SPIE.9147E..2VS} , DESI \citep[]{2015AAS...22533607S}, 4MOST\footnote{https://www.4most.eu/cms/} and Euclid\footnote{http://sci.esa.int/euclid/} \citep[]{2011arXiv1110.3193L, 2015arXiv150502165S} will use emission-line galaxies (ELGs) as BAO tracers to explore the Universe large-scale structure out to $z\sim 2$. Observing ELGs, learning how to model their clustering properties and understanding how they populate their host halos are therefore crucial points that we need to understand in order to select the targets for future experiments. From the observational point of view, the recent increment of available ELG spectroscopic data \citep{2014A&A...566A.108G,2015A&A...575A..40C} allows one to measure their clustering properties over about 12 deg$^2$ at $z=0.8$ (corresponding to a comoving volume of $V\sim10.6\times 10^6\,h^{-3}$Mpc$^3$ in the Planck cosmology; see Section \ref{sec:clustering} for details), which represents a dramatic improvement. 

\cite{2013MNRAS.433.1146C} demonstrated that neither a standard HOD nor a traditional SHAM technique are able to reproduce the angular clustering of ELGs on small scales. In fact, both techniques are based on the assumption that the galaxy sample to model is complete, but this is not the case of the ELGs, which are highly incomplete in stellar mass. One could instead use semi-analytic models of galaxy formation and hydrodynamic simulations, but they lack of mass resolutions to model emission line galaxies.

The aim of this work is to provide a modified version of the standard SHAM prescription, directly based on the latest MultiDark N-body simulation with Planck cosmology, that accounts for the ELG incompleteness and returns suitable mock galaxy catalogs able to accurately predict the ELG angular and redshift-space clustering, respectively, on small and larger scales. These mock catalogs are released to the public. 

The paper is organized as follows. Section \ref{sec:dataall} describes the data sets and the MultiDark simulation box used in our analysis. In Section \ref{sec:clustering} we present our ELG clustering and weak lensing measurements. In Section \ref{sec:SHAM_procedure} we explain how we model the ELG clustering and we present our main results. Section \ref{sec:clusteringTrends} discusses the implications of our ELG clustering analysis in a galaxy evolution perspective, and Section \ref{sec:summary} summarizes our main results.

Throughout the paper, we assume the Planck cosmology \citep{Planck2014} and magnitudes in the AB system \citep{1983ApJ...266..713O}.


\section{Data and Simulation}
\label{sec:dataall}

\subsection{Data sets}
\label{sec:data}
We build our ELG galaxy sample using the Canada-France-Hawaii Telescope Legacy Survey (CFHT-LS) Wide T0007\footnote{http://www.cfht.hawaii.edu/Science/CFHTLS/} photometric redshift catalog \citep{Ilbert_06,Coupon_2009}. We apply a $g$-band magnitude cut, $20<g<22.8$ \citep{1996AJ....111.1748F}, to select galaxies with bright emission lines and low dust at $z<1$. We also apply a color selection, $-0.5<(u-r)<0.7\,(g-i)+0.1$, to remove the low-redshift galaxies. For details on the selection function, see \cite{2015A&A...575A..40C}.
Then, to obtain the largest possible area, we convert the $i$-selection into the new Megacam $i$-band filter\footnote{\url{http://www4.cadc-ccda.hia-iha.nrc-cnrc.gc.ca/en/megapipe/docs/filt.html}}. 
For the W1, W3 and W4\footnote{\url{http://www.cfht.hawaii.edu/Science/CFHLS/T0007/T0007-docsu10.html}} fields, we derive an average density of about 500 ELGs per  deg$^{2}$, 70\% of which have a photometric redshift in the range $0.6<z<1$. The densities of each field are reported in Table \ref{table:photoELG}, and the errors on the photometric redshift are $\sigma_z<0.05\,(1+z)$ for $i<22.5$ and $z<1$. The $ugri$ ELG selection is brighter than $i<22.5$.

\begin{table*}
\caption{ELG photometric data per CFHT-LS Wide field after applying the bright star and bad field mask.}
\begin{center}
\begin{tabular}{l c c c c c c}
\hline  
Field & W1 & W3 & W4 & all \\ 
Center $\alpha,\delta$ & 35$^\circ$, -7$^\circ$ & 215$^\circ$, 54$^\circ$ & 333$^\circ$, 2$^\circ$ & -\\ 
area [deg$^{2}$]  &  63.75 & 44.22 & 23.3 &  131.27  \\ \hline
N &  32,808 & 22,195 & 11,025 &  66,028  \\
N [deg$^{-2}$] &  514.64 & 501.92 & 473.18 &  502.99  \\
$z_{phot}$ quartiles &  0.78 / 0.88 / 1.03   & 0.77 / 0.88 / 1.05   & 0.78 / 0.88 / 1.03   & 0.78 / 0.88 / 1.03    \\
$z_{phot}$ Deciles D10, D90 &  0.7 / 1.24   & 0.68 / 1.29   & 0.69 / 1.25   & 0.69 / 1.26    \\
\hdashline
N ($0.6<z_{phot}<1$) & 23,433 & 15,242 & 7,861 &  46,536  \\
N ($0.6<z_{phot}<1$) [deg$^{-2}$] & 367.58 & 344.69 & 337.38 &  354.51  \\
$z_{phot}$ quartiles &  0.75 / 0.83 / 0.9   & 0.74 / 0.81 / 0.89   & 0.75 / 0.83 / 0.89   & 0.75 / 0.82 / 0.89   \\
\hline
\end{tabular}
\end{center}
\label{table:photoELG}
\end{table*}

\begin{figure*}
\begin{center}
\includegraphics[width=6.1cm]{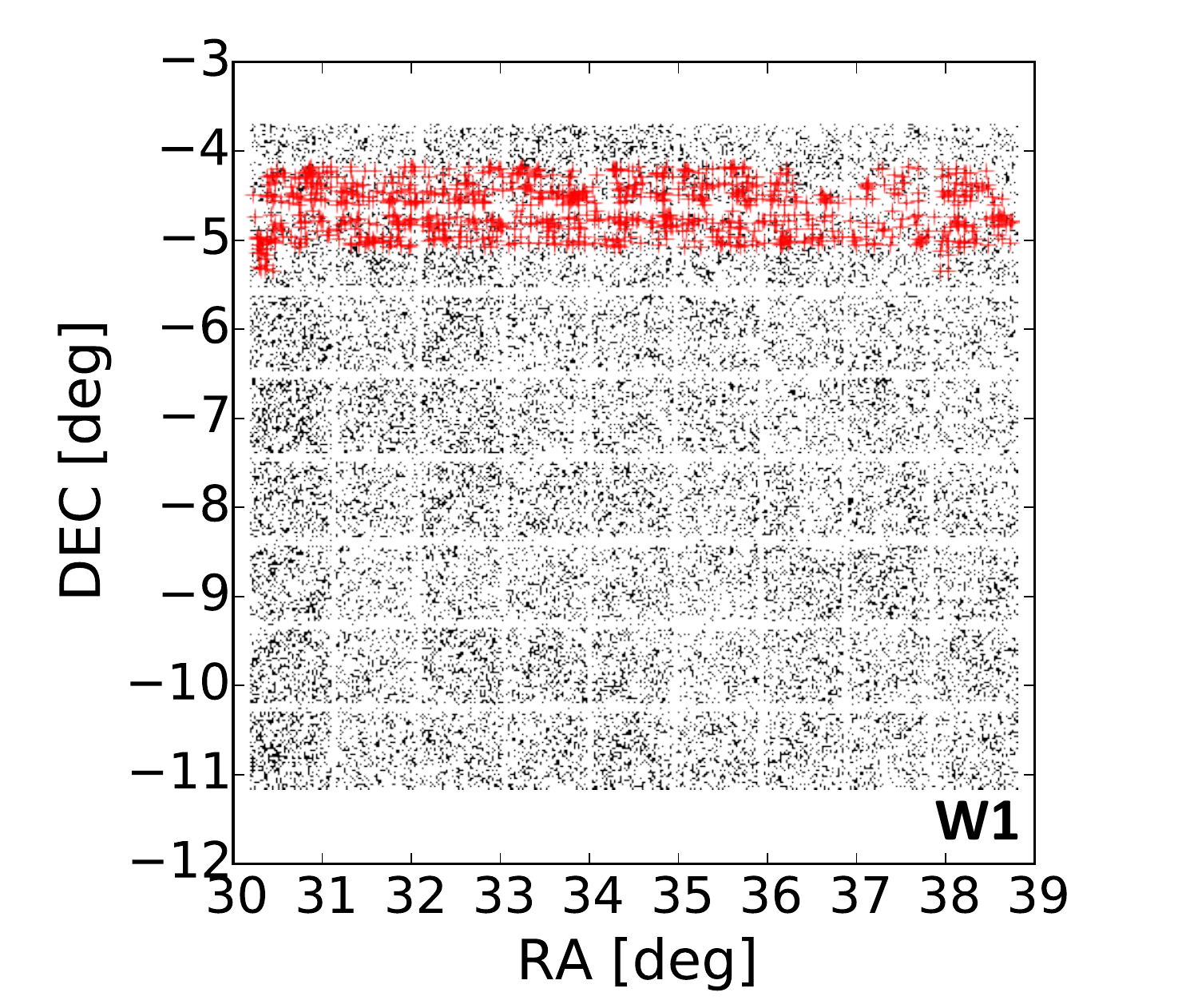}\hfill
\includegraphics[width=5.8cm]{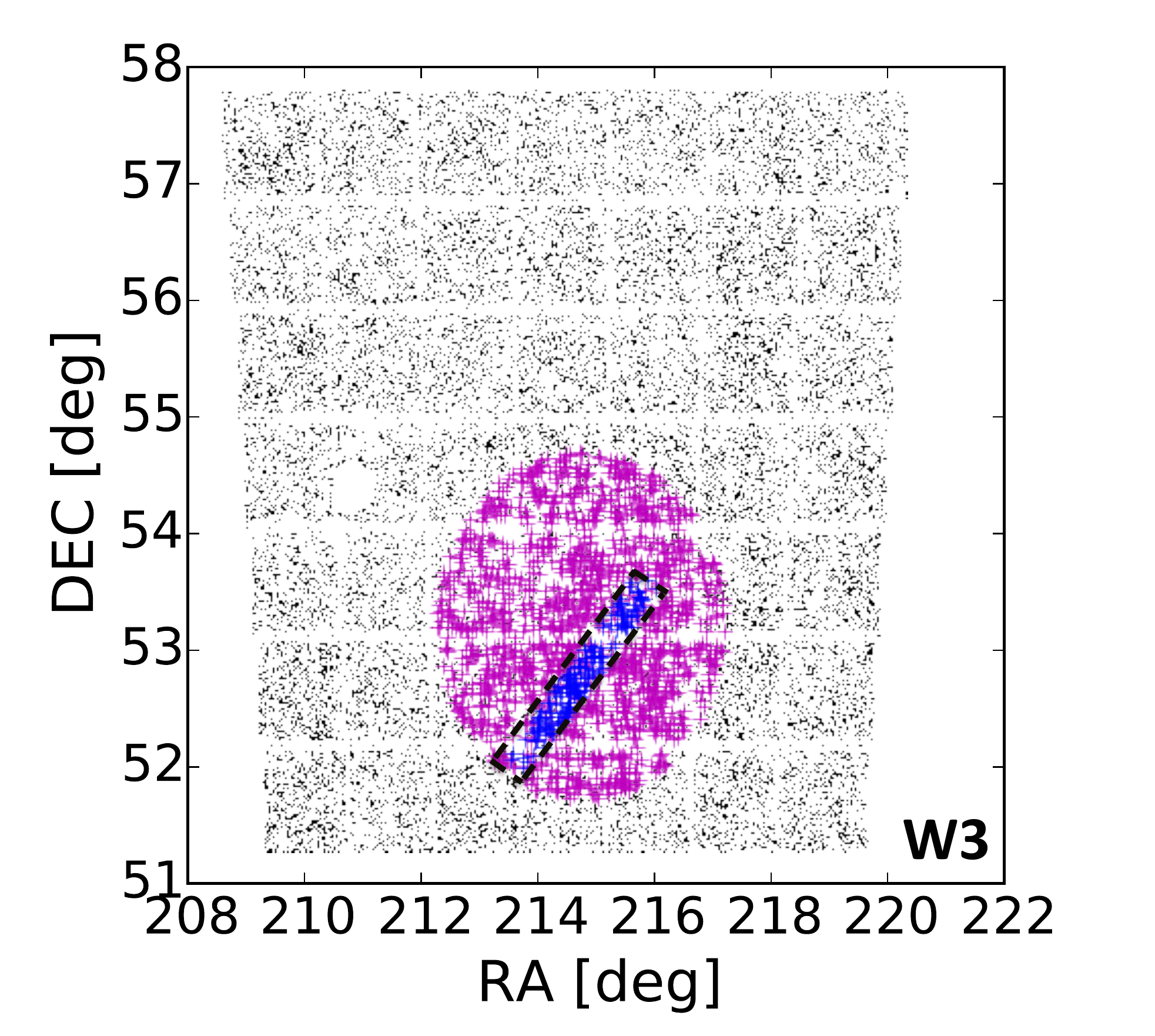}\hfill
\includegraphics[width=5.8cm]{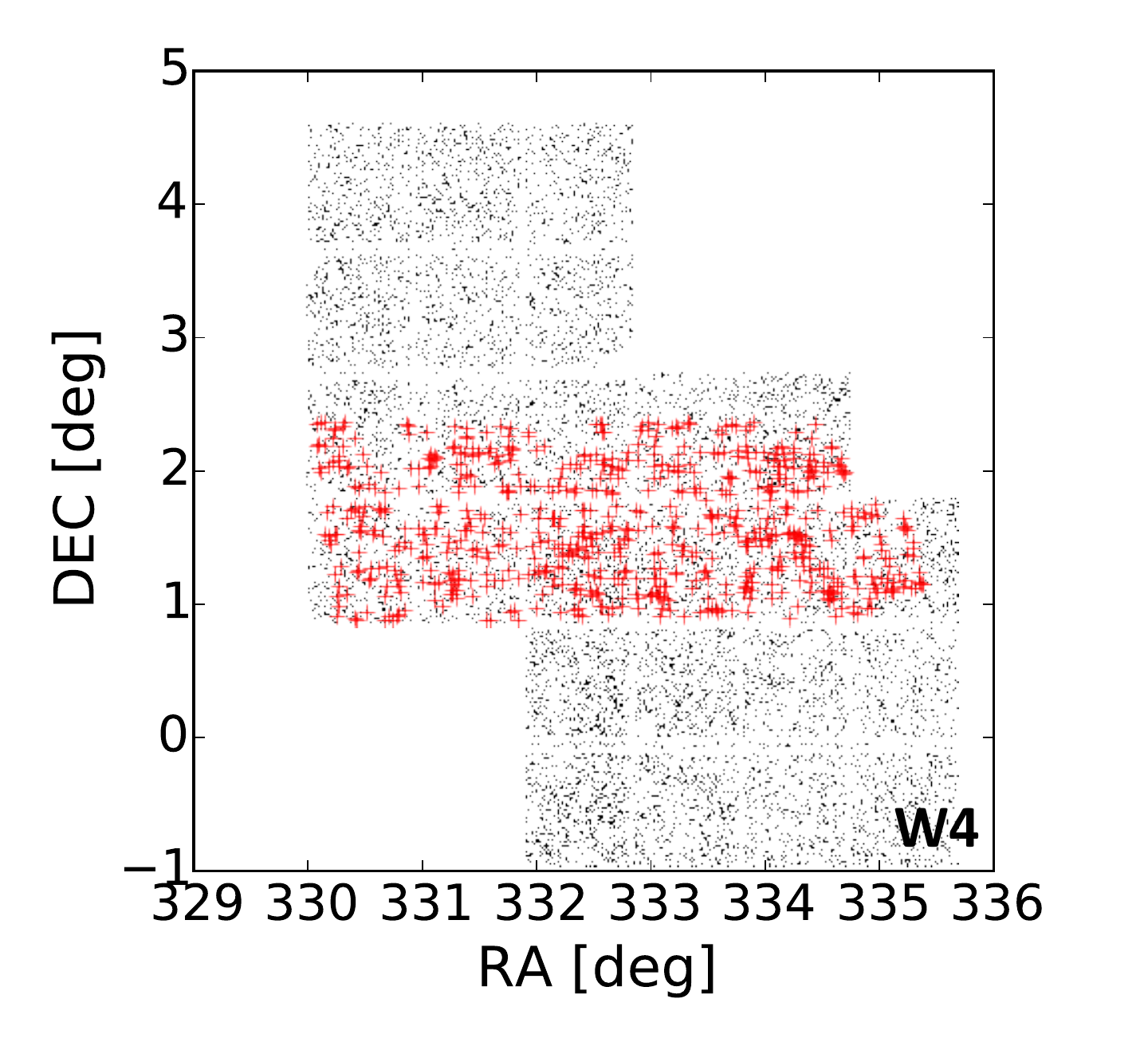} \hfill
\caption{Photometric (black points) and spectroscopic (VIPERS: red crosses in the right and left panels; BOSS: magenta crosses forming the oval in the central panel; DEEP2: blue crosses in the dashed box in the central panel) coordinates of our ELG sample in the three CFHT-LS Wide fields.}
\label{fig:ra:dec}
\end{center}
\end{figure*}

\begin{figure}
\begin{center}
\includegraphics[width=8cm]{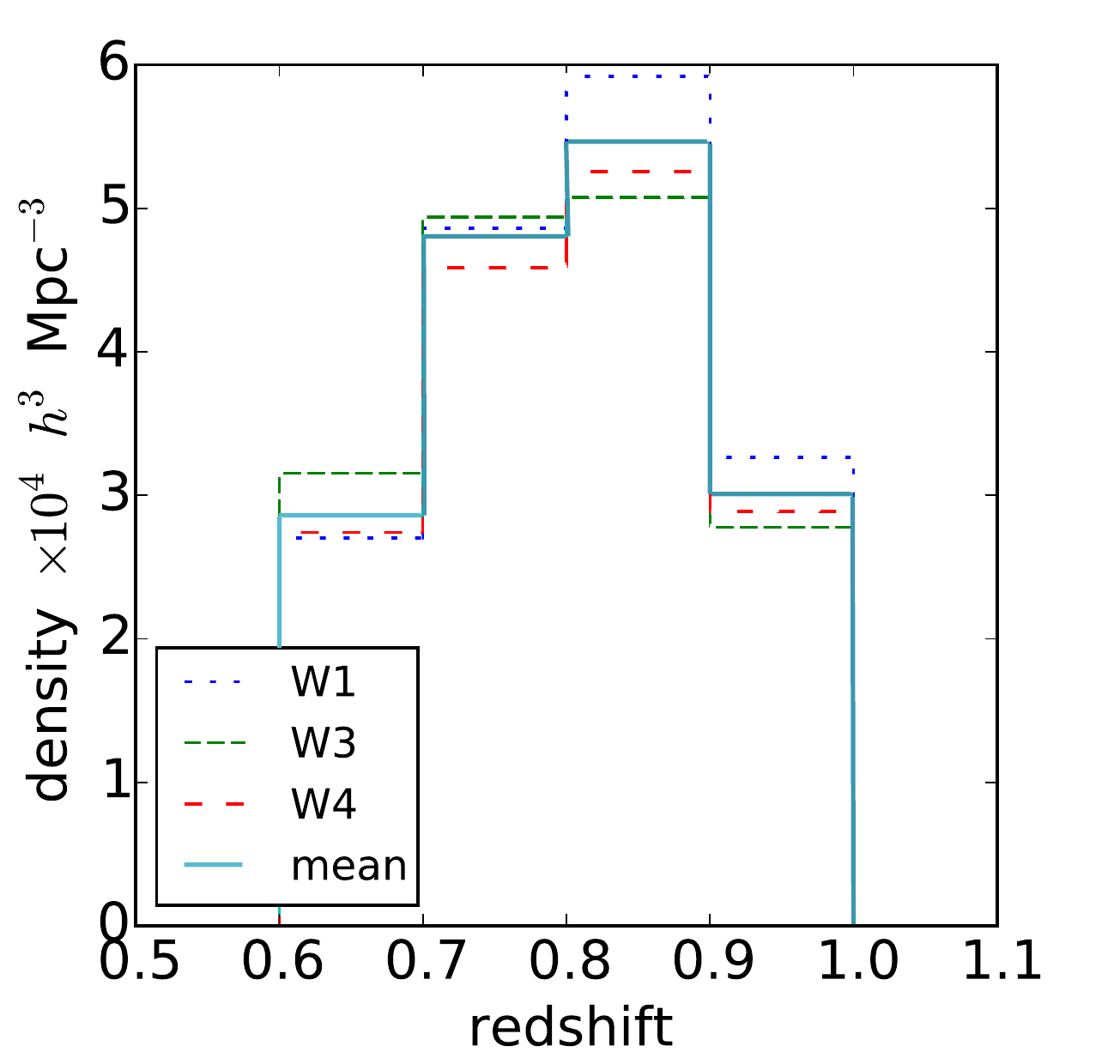}\hfill
\caption{ELG weighted spectroscopic redshift distribution per unit volume for the W1, W3, and W4 Wide fields (dashed and dotted histograms), and their mean value (solid line).}
\label{fig:nz}
\end{center}
\end{figure}

We match the photometric targets to the available spectroscopic surveys -- BOSS DR12, DEEP2, VIPERS \citep{2012AJ....144..144B,2015ApJS..219...12A,2013ApJS..208....5N,2014A&A...566A.108G} -- within $1"$ radius; see Table \ref{table:spectroELG}. Based on KS-tests, the VIPERS, BOSS and DEEP2 spectroscopic selections constitute fair sub-samples of the complete selection: the hypothesis that they are drawn from the same distribution cannot be rejected at the 90\% confidence level. For these samples, we create random catalogs with the same redshift distribution of the data and 30 times denser. Figure \ref{fig:nz} displays the ELG spectroscopic redshift distribution per unit volume for the three Wide fields (dashed, dotted, dot-dashed histograms), and their mean (solid line). Two thirds of the galaxy density is located in the redshift range $0.7<z<0.9$, while both the intervals $0.6<z<0.7$ and $0.9<z<1$ contain one sixth of the sample. According to the ELG selection function in \cite{2015A&A...575A..40C}, we select only galaxies at $z>0.6$ since we are not interested in low-redshift objects. We have investigated further the impact of the higher redshift cut, $z<1$, on the angular clustering by imposing to the ELG sample different redshift thresholds: $z<1, 1.2, 1.4, 1.6$. In all these samples the lower redshift cut is fixed at $z>0.6$ and we have imposed the $i<22.5$ magnitude cut to eliminate bad photometric redshifts.
We find that including also ELGs at $z\ge1$, we are slightly enhancing the galaxy number density of our sample and consequently suppressing the amplitude of $w(\theta)$, but we do no see any substantial change in the angular clustering trend with respect to the $z<1$ case. We therefore restrict the analysis to the redshift range $0.6<z<1$.

\begin{table*}
\caption{ELG spectroscopic data. }
\begin{center}
\begin{tabular}{l r r r r r r r r r r r r r r r}
\hline  
survey & match & good $z$ & $0.6<z<1$ & area [deg$^2$]  & $\bar{z}$\\ \hline
VIPERS W1 	& 1,223 	& 942 	& 760 	& 5.478 &0.803	\\ 
  \hdashline
BOSS W3 	&  2,145 	& 1,876 	&1,357	& 6.67 & 0.803  \\  
DEEP2 W3 	&225 	& 222 	& 156 	&0.5 &0.803\\ \hdashline
VIPERS W4	&  1,148 	& 846	& 680 	& 5.120 & 0.795 \\ \hdashline
All 			& 4,741	& 3,886 	& 2,953	& 17.668 & 0.803 \\
\hline
\end{tabular}
\end{center}
\label{table:spectroELG}
\end{table*}%


\subsection{MultiDark simulations}
\label{sec:MD_simulation}

The MultiDark Planck simulation (MDPL, \citealt{2014arXiv1411.4001K}; \url{www.MultiDark.org}) contains $3840^3$ particles in a 1$\,h^{-1}$Gpc box, and was created adopting Planck $\Lambda$CDM cosmology \citep{Planck2014}. Halos are identified based on density peaks including substructures using the Bound Density Maximum (BDM) halo finder \citep{1997astro.ph.12217K, Riebe2013}.

We use the MDPL halo catalogs to build a mock light-cone that matches the mean ELG redshift distribution shown in Figure \ref{fig:nz}. Given the high density of the ELG tracers and their expected low-mass host halos, the MDPL box is an excellent compromise between numerical resolution and volume. We apply the SUrvey GenerAtoR code (SUGAR; Rodriguez-Torres et al. (2015), in prep.) to the 11 snapshots available from MDPL to construct a light-cone with a volume ten times the observations that covers the redshift range 0.6$<z<$1 ($\sim 1 h^{-1}$ Gpc depth). The procedure used is analogous to the method presented by \cite{2005MNRAS.360..159B} and \cite{2007MNRAS.376....2K}, and can be summarized as follows:

\begin{enumerate}
\item Set the properties of the light-cone: angular mask, radial selection function (number density) and number of snapshots within the redshift range considered. Each slice of the light-cone is constructed by selecting all halos from every MDPL snapshot. The thickness of a slice at redshift $z_i$ is given by $[(z_i+z_{i-1})/2,(z_i+z_{i+1})/2]$\\
\item Place an observer (i.e., $z=0$) inside the box and shift the cartesian coordinates of the box in such a way that the observer occupies the central point of the box at $z=0.8$\\
\item Convert from cartesian $(x,y,z)$ to spherical $(\alpha, \delta, r_c)$ coordinates, where $r_c$ is the comoving distance in real space. The redshift of each point will be:
\begin{equation}
r_c(z)=\int^b_a\frac{cdz'}{H_0\sqrt{\Omega_m(1+z')^3+\Omega_{\Lambda}}}
\end{equation}\\
\item From each snapshot, select the (sub)halos so that $(z_i+z_{i-1})/2<z<(z_i+z_{i+1})/2$ and $\alpha/\delta$ lie inside the sky window. Since the ELG observational data represent halos with typical masses $\sim10^{12}h^{-1}$M$_{\odot}$, in the light-cone we include all halos for which the simulation is complete i.e., $\log (M_h/h^{-1}$M$_{\odot})>11.2$\\
\item Using the halo velocities, $v_p$, we compute the peculiar velocity contribution for each object along the line-of-sight and derive its distance in redshift-space as
\begin{equation}
s=r_c+(v_p\cdot r_c)/(aH(z)),
\end{equation}
where $a=(1+z)^{-1}$ is the scale factor and $H(z)$ is the Hubble parameter at redshift $z$\\
\item Finally, select objects from the light-cone using our selection function.
\end{enumerate}

Throughout the paper we will designate our lightcone as ``MDPL-LC''.
Section \ref{sec:SHAM_procedure} describes in detail the halo selection and the (Sub)Halo-Abundance Matching modeling adopted to determine the halo occupation distribution of our ELG sample.


\section{Measurements}
\label{sec:clustering}

Using the ELG sample described in Section \ref{sec:data}, we measure both galaxy clustering and galaxy-galaxy lensing. The following provides a detailed description of our measurements.

\subsection{Galaxy Clustering}
\label{sec:gal_clustering}

\begin{figure}
\begin{center}
\includegraphics[width=8cm]{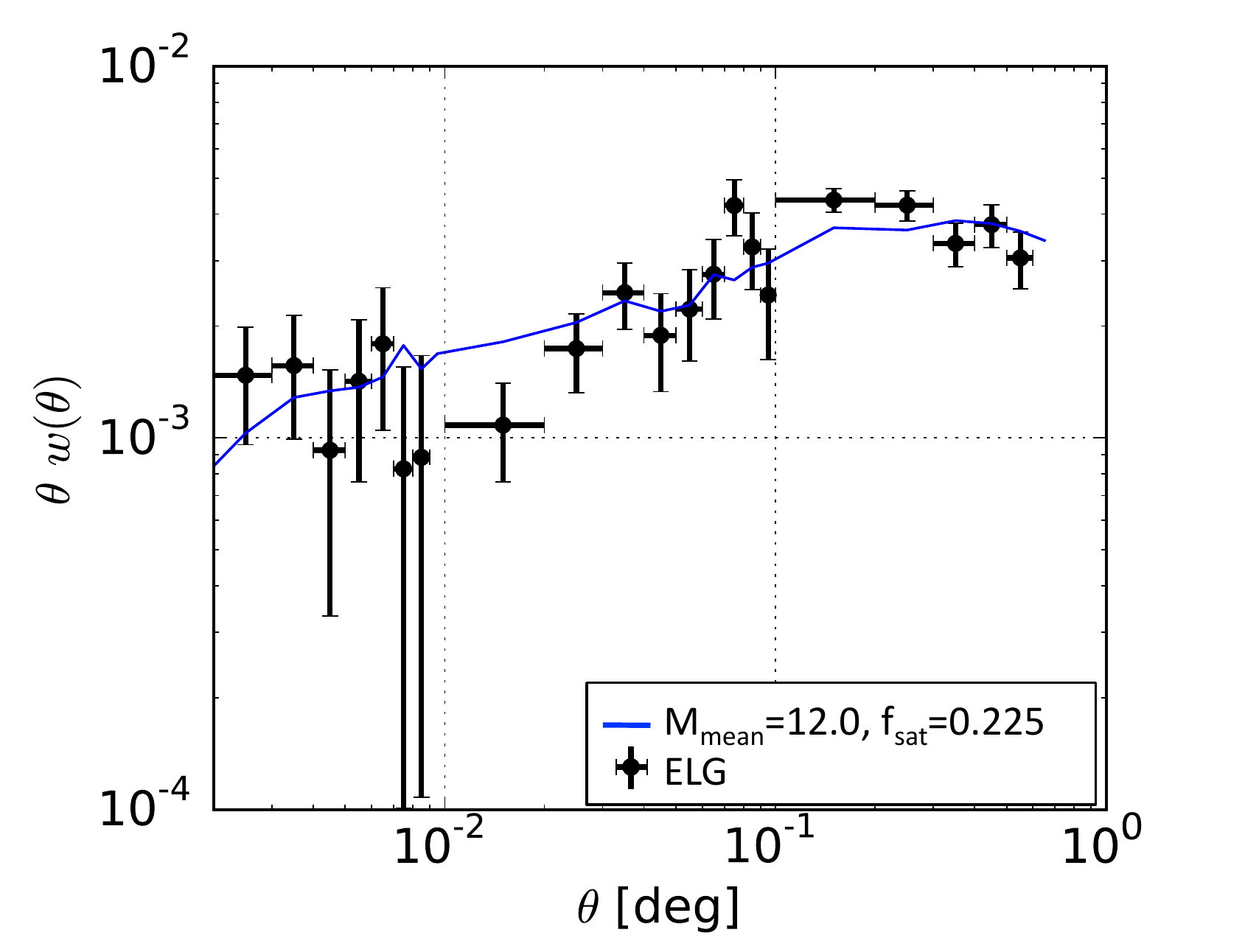}\\
\includegraphics[width=8cm]{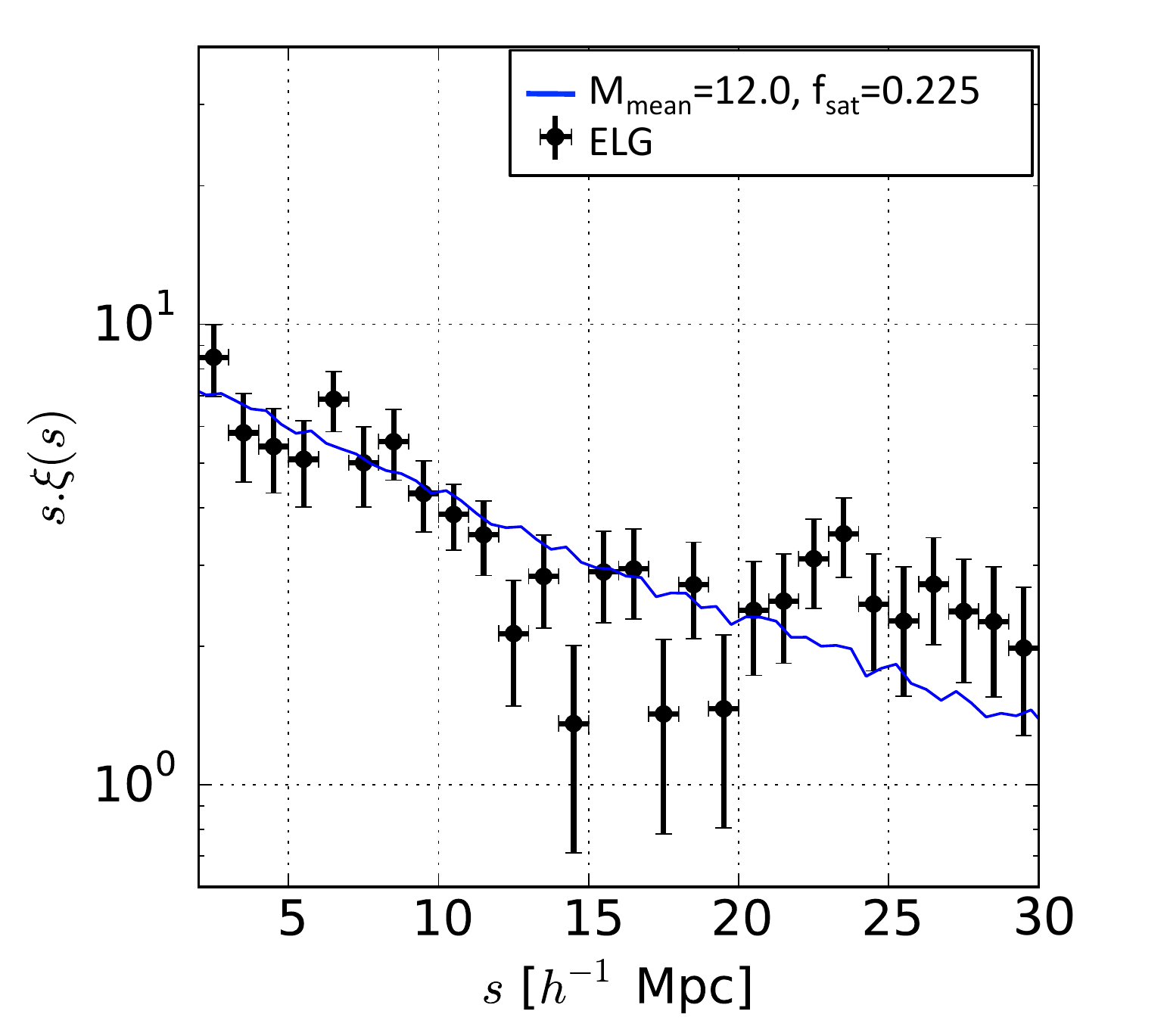}
\caption{Two-point angular (top panel) and redshift-space (bottom panel) ELG correlation functions (points), together with our best-fit model (blue line), which corresponds to the point highlighted by a star in Figure \ref{clustering:plot:best:model}.}
\label{clustering:measurement:plot}
\end{center}
\end{figure}

We estimate both the angular, $w(\theta)$, and the redshift-space, $\xi(s)$ (hereafter $\xi^s$), two-point correlation functions following the procedures described by \cite{1993ApJ...412...64L}, \cite{2012A&A...542A...5C} and \cite{2013A&A...557A..54D}. 

To compute $\xi^s$ on the VIPERS and the BOSS ELG samples (see Table \ref{table:spectroELG}), independently, we create linear bins in separations of $1\,h^{-1}$Mpc at $s<10\,h^{-1}$Mpc, and $4\,h^{-1}$Mpc for $10<s<40\,h^{-1}$Mpc. 
We then correct the impact of redshift errors and catastrophic redshifts to recover the correlation function down to $1\,h^{-1}$Mpc. The ELG we are targeting are observed using three plates overlapping the same area of the sky. This configuration guarantees that all the targets are observed at the end of the process and there is no fiber collision \citep[]{2003AJ....125.2276B, 2014MNRAS.444..476R}. We correct for the finite size of VIPERS following \citet{2013A&A...557A..54D}, and define the survey completeness in terms of target sampling rate (TSR) and the spectroscopic success rates (SSR). The first quantity weights our ability of obtaining spectra from the potential targets meeting the survey selection in the parent photometric sample. For each galaxy in the spectroscopic catalog, we count the number of objects that lie within a given radius in the spectroscopic ($N_{targeted}$) and in the photometric ($N_{parent}$) sample. The TSR is then given by the ratio $TSR=N_{targeted}/N_{parent}$. The SSR represents our ability of determining galaxy redshifts from observed spectra. To compute it, we replace the bad redshifts in the spectroscopic catalog with good photometric redshifts. Galaxies with good spectra are assigned flag=0 ($N_{good}$); galaxies with replaced redshifts are assigned flag=1 ($N_{targeted}$). The SSR is then computed as the ratio $SSR=N_{good}/N_{targeted}$. The contribution of each galaxy in the clustering pair counts is then weighted by $w=1/(TSR*SSR)$. Finally, we combine the VIPERS and BOSS measurements weighted by the projected density of each field.
The resulting redshift-space correlation function is displayed in the bottom panel of Figure \ref{clustering:measurement:plot}; by fitting a power-law model, $\xi(s)=(s/s_0)^{\alpha}$, in the separation range $2<s<30\,h^{-1}$Mpc, we find $s_0=(5.3\pm0.2)\,h^{-1}$Mpc and $\alpha={-1.6\pm0.1}$.

Analogously, we calculate the angular 2PCF, $w(\theta)$, using photometric redshifts from the W1, W3 and W4 CFHT-LS fields. The points are corrected from the integral constraint following \cite{2010ApJ...724..878T} and \cite{2012A&A...542A...5C} to account for the restricted area of observation. On scales $\theta<0.05^\circ$, all three fields provide consistent measurements. At larger scales, the clustering signals in the W1 and W4 fields do not decrease as rapidly as expected, probably pointing to possible systematics that need to be investigated further. We therefore use only the measurement on the W3 field, which appears the most robust (see Figure \ref{clustering:measurement:plot}, top panel).
The $w(\theta)$ of the W3 field is in perfect agreement with Figure 9 (panel 4) in \citet{2013MNRAS.433.1146C}. This result was computed on the Stripe 82 region \citep{2002AJ....123..485S}, with three times larger area.
At the mean redshift of the sample, $z=0.8$, one degree corresponds to 18.847$\,h^{-1}$Mpc; thus, $w(\theta)$ spans the range from $\sim40\,h^{-1}$kpc up to $\sim20\,h^{-1}$Mpc. 
We investigate further the impact on the clustering amplitude of including the tails of the photometric distribution, i.e. $z_{phot}<0.6$ and $z_{phot}>1$. This inclusion does not produce any substantial change in $w(\theta)$, except for some additional noise. We also test how photometric uncertainties affect the clustering errors computed via mock resampling. To this purpose, following \cite{2012A&A...542A...5C}, we perturb our original redshift distribution by applying a photometric scatter with mean $\sigma_z=0.035(1+z)$. We then quantify the number of photometric objects that, due to this scatter, enter the ELG selection in the range $0.6<z<1$ from the lower and higher tails of the distribution, and the objects that exit. We find that only $2.5\%$ photometric redshifts enter the ELG selection in the range $0.6<z<1$ from the upper and lower tails, and their effect on the clustering is negligible.

To estimate the errors on our galaxy clustering measurements, since the simulated light-cone area is larger than the data ($\sim$560 deg$^2$), we divide the best MDPL-LC model into independent (i.e., non-overlapping) realizations of our ELG data (8 for the photometric and 24 for the spectroscopic samples), and obtain sample variance diagonal errors that we use rather than Poisson errors. Including the photometric uncertainties in our jackknife resamplings does not provoke any significant change in the error estimates. We neglect a full-covariance analysis because the number of sub-samples we have is too small to produce reliable covariance estimates. Including also the off-diagonal elements of the covariance matrices would result in large fluctuations of the error bars. Of course, excluding covariances we are adopting a simplified approach, but it provides a good sense of how the SDSS BOSS ELG clustering behaves. On the other hand, the ELG sample considered here is too sparse to derive tight constraints from our clustering analysis. New-generation large-volume spectroscopic surveys as eBOSS, DESI and 4MOST, will provide new data with unprecedented statistics, sky coverage and imaging quality. Using those data, a fully covariant approach will return reliable and accurate error estimates.

We compare the combined $\xi^s$ measurement from BOSS and VIPERS to previous measurements by \citet{2013A&A...557A..17M} to provide a first interpretation. Our result matches both the clustering signal of galaxies selected in the stellar mass range $9.5<\log \; ( M_*/h^{-1}$M$_\odot ) <11$, and the clustering of galaxies selected by absolute magnitude in the interval $-22<M_B - 5 \log(h)<-20.5$.
Using the stellar-to-halo mass relation from \cite{2012ApJ...744..159L}, \cite{2015arXiv150200313S} and \cite{2015MNRAS.449.1352C}, we can deduce a rough estimate of the halo masses populated by our ELG sample i.e., $11.6<\log \; ( M_h/h^{-1}$M$_\odot ) <12.7$. These halo masses are typical of Milky-Way size halos, being much less massive than those hosting the LRG sample, see \cite{2013MNRAS.432..743N}.

In the angular clustering measurement, the change of slope occurs at $\theta\sim0.01^\circ$, corresponding to $\sim$ 200 $h^{-1}$kpc. Using MDPL, we derive the relation between halo mass and virial radius at $z\sim 0.8$; halos with virial radius $\sim200 h^{-1}$kpc occupy the mass range M$_h=(0.5-1)\,\times10^{12}h^{-1}$M$_\odot$. Since a single galaxy per halo would not induce such a change in the $w(\theta)$ slope, this result implies a satellite fraction of approximately $22.5\%$ (see Section \ref{sec:SHAM_procedure}). Figure \ref{clustering:measurement:plot} displays a good agreement between our clustering measurements and predictions for ELG halos of mass $10^{12}h^{-1}$M$_{\odot}$ with this satellite fraction.


\subsection{Weak lensing}
\label{sec:weak_lensing}

We use the latest weak lensing catalogs produced by the Canada-France-Hawaii Telescope Lensing Survey \citep[CFHTLenS;][]{2012MNRAS.427..146H,2013MNRAS.433.2545E} on the W1 and W3 fields to measure the galaxy-galaxy lensing around 47,485 ELG lenses. This measurement allows one to constrain the halo masses. We follow \citet{2013MNRAS.431.1439G} and apply only the multiplicative correction, $m_s$, to the shear measurement and avoid the c2 correction.
We measure the tangential shear, $\gamma^t$, around the photometric ELG sample as a function of the radial distance from the lenses using the \citep{2011MNRAS.412.2665V} estimator: 
\begin{equation}
\Delta \Sigma = \left[ \frac{\sum_{ls} w_{ls} \gamma^t_{ls}  \Sigma_c} {\sum_{ls} w_{ls}} \right] /  \left[ \frac{\sum_{ls} w_{ls} (1+m_s)}{\sum_{ls} w_{ls} } \right],
\end{equation}
where the sum runs over the lens - source pairs ($ls$) and the $w_{ls}$ values are the weight obtained by lensfit.

Since the lenses are at the higher tail of the redshift distribution and the ELGs are expected to live in low-mass halos, we recover a low signal-to-noise ratio around 2 for $R<1$ Mpc.

\begin{figure}
\begin{center}
\includegraphics[width=8cm]{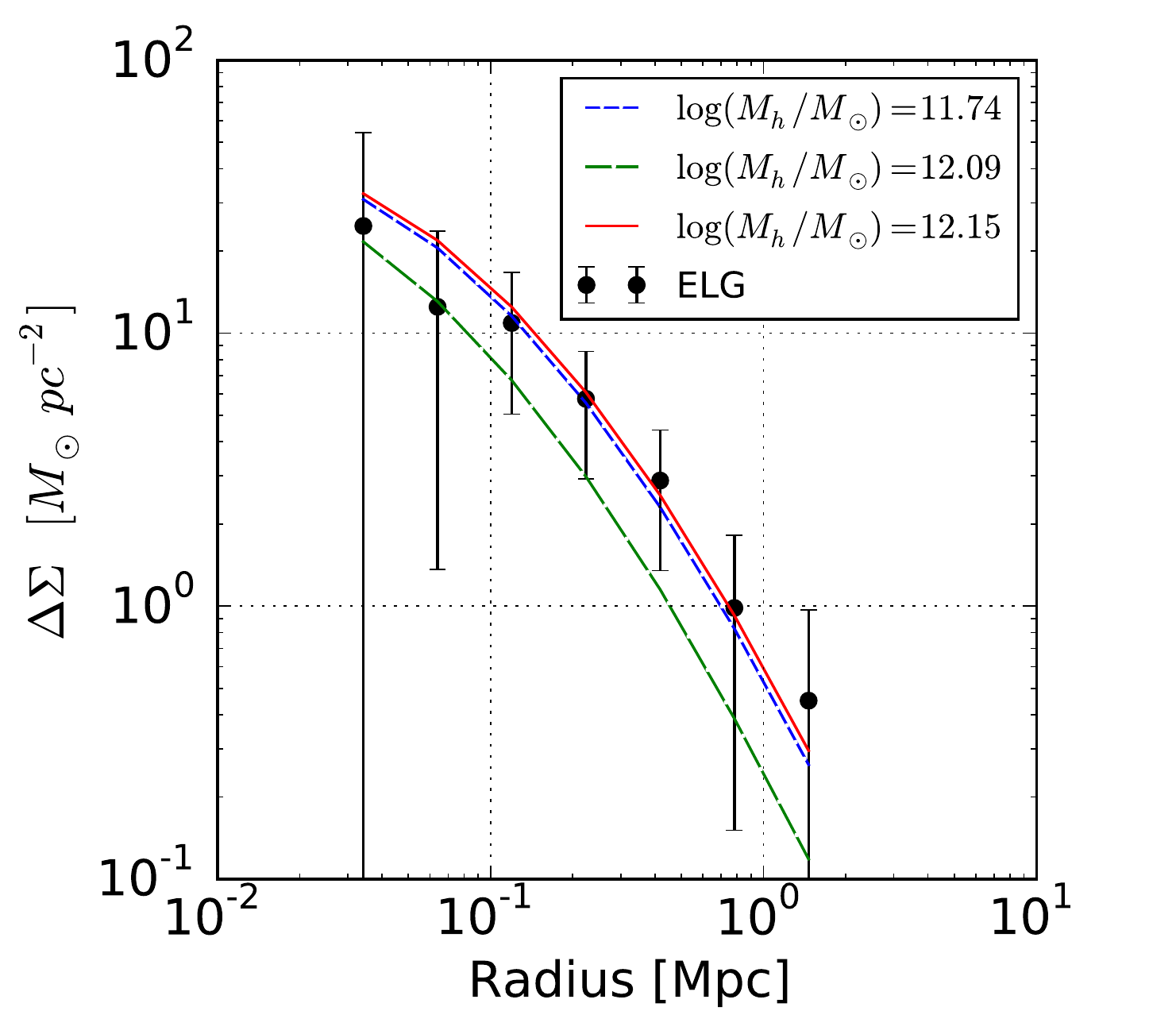}
\caption{ELG surface density ($\Delta \Sigma$) as a function of the physical scale for different lens models.}
\label{lensing:plot}
\end{center}
\end{figure}

We model the measurement using a truncated Navarro, Frank $\&$ White (NFW) halo profile \citep{2009JCAP...01..015B} and the mass-concentration relation from \citet{2007MNRAS.381.1450N} to truncate halos at half their concentration \citep{2000ApJ...534...34W}. 
The best-fit model suggests typical halo masses of $M_{200}=1.25\pm0.45 \times 10^{12}  h^{-1}M_\odot$. The lower and upper mass limits are, respectively, $M_{200}=5.61\pm7.20 \times 10^{11}  h^{-1}M_\odot$ and $M_{200}=1.41\pm0.51 \times 10^{12} h^{-1}M_\odot$; see Fig. \ref{lensing:plot}.
This measurement is in good agreement with the first interpretations based on the clustering (see Section \ref{sec:clustering}).
 
\section{Halo Occupation for emission line galaxies}
\label{sec:SHAM_procedure}

The (Sub)Halo-Abundance Matching \citep[SHAM; e.g.,][]{Conroy2006, Trujillo2011} technique is a straightforward method to link observed galaxies with dark-matter-only simulated halos. It relies in a monotonic correspondence between halo and galaxy number densities, which is based on the assumption that more luminous galaxies reside in more massive halos. Such association is performed by choosing suitable proxies for both halos and galaxies (e.g., the halo maximum circular velocity and the galaxy luminosity or stellar mass) and includes some scatter (see \cite{Trujillo2011} for details). 
The advantage of using $N$-body simulations, compared to analytical models, is given by the accuracy achieved in the predictions of the clustering for a given halo population. Many state-of-the art clustering measurements have been modeled using a SHAM technique that maps the observations onto suitable high-resolution N-body simulations, allowing the interpretation of the halo occupation distribution and bias \citep{2013A&A...557A..54D,2013MNRAS.432..743N, 2015MNRAS.447..646C}. 
\cite{2015MNRAS.446..651W} recently presented a method to upgrade SHAM models to account for differences between quenched and star-forming galaxies. 

In the specific case of the emission-line galaxies, the traditional SHAM approach cannot be applied since it requires a complete galaxy sample, and ELGs are far from being complete in any parameter space, even in terms of their emission line luminosity, see \cite{2013MNRAS.433.1146C}. We therefore must modify the standard SHAM procedure to take into account the ELG incompleteness and match their clustering amplitude. 
To this purpose, we selected halos and subhalos by mass (for the subhalos we considered only the mass of the bound particles, to avoid ambiguities) to be able to compare directly with the weak lensing measurements. In the future, provided a high signal-to-noise ratio in the clustering measurement, we will properly select (sub)halos by their maximum circular velocity at accretion, \citep[e.g., ][]{2013ApJ...763...18B}.

In order to model both the $1$-halo and the $2$-halo terms in the ELG two-point correlation functions and the weak lensing measurement, we use the MultiDark Planck 1$h^{-3}$Gpc$^3$ box (see Section \ref{sec:MD_simulation}), which represents the best compromise between high resolution and volume, as previously described in Section \ref{sec:clustering}. 

\begin{figure*}
\begin{center}
\hspace{-0.3cm}
\includegraphics[width=7.5cm]{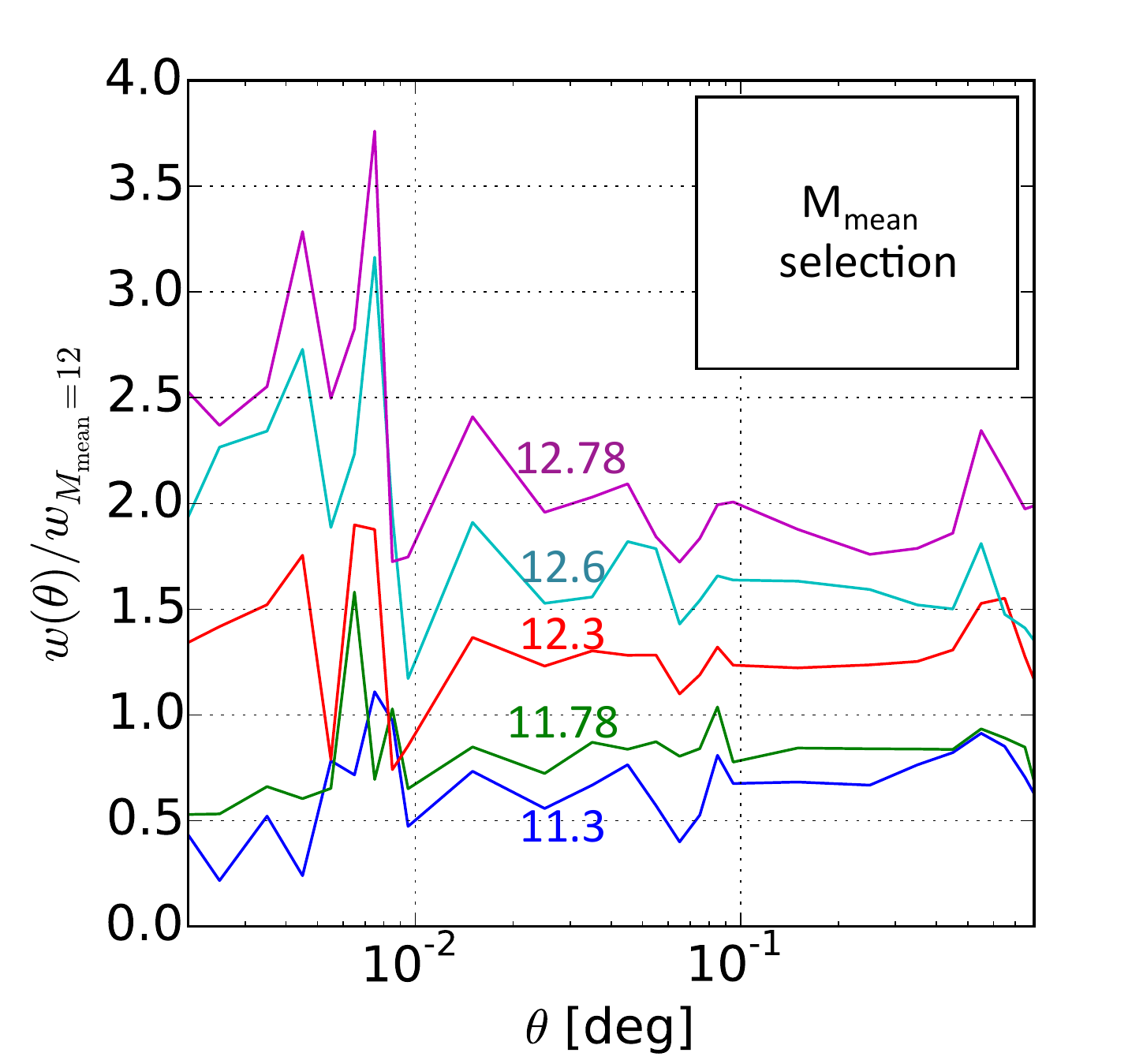}\hfill
\hspace{1cm}
\includegraphics[width=7.35cm]{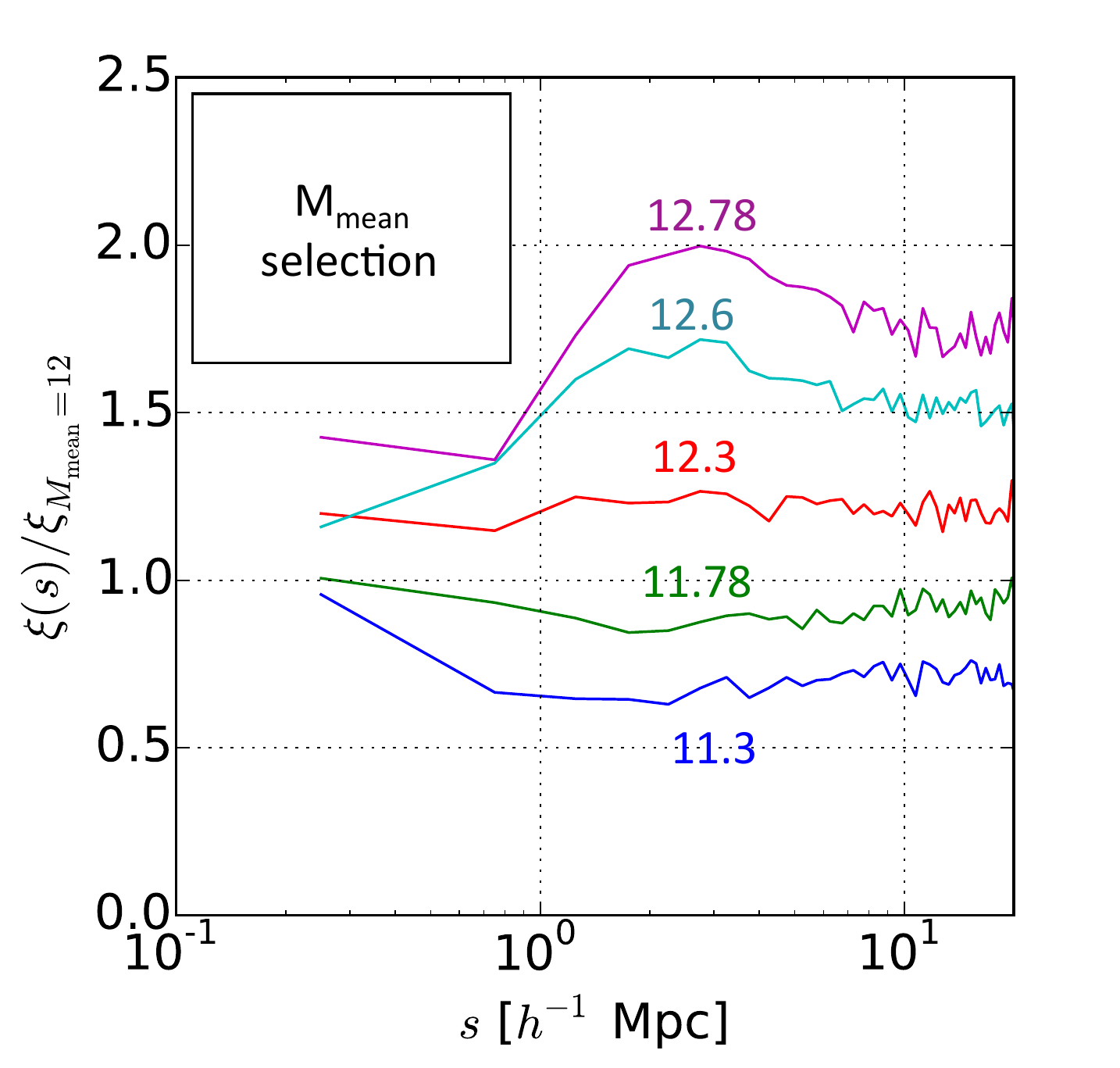}\hfill\\
\hspace{-0.3cm}
\includegraphics[width=7.4cm]{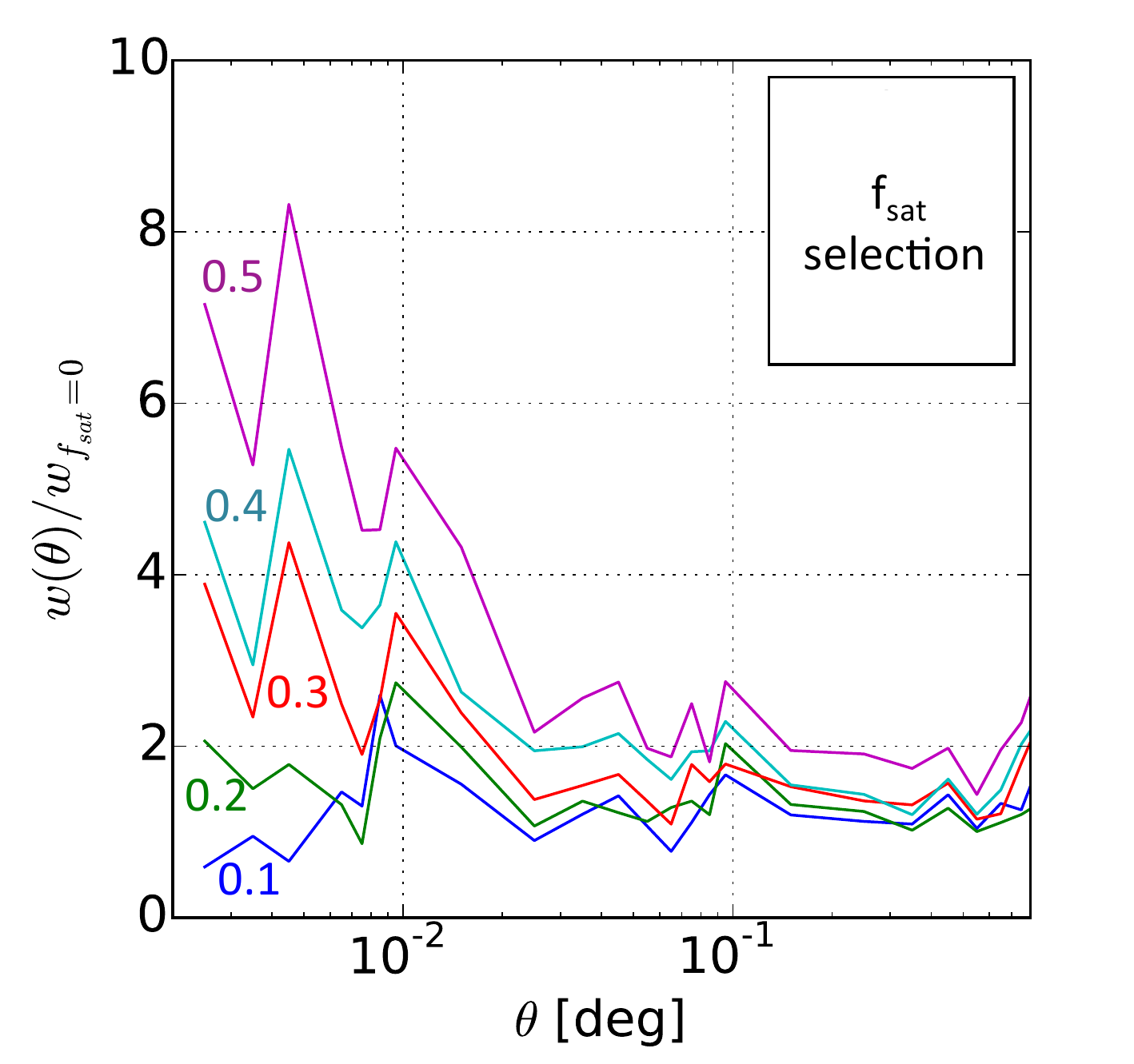}\hfill
\hspace{1cm}
\includegraphics[width=7.35cm]{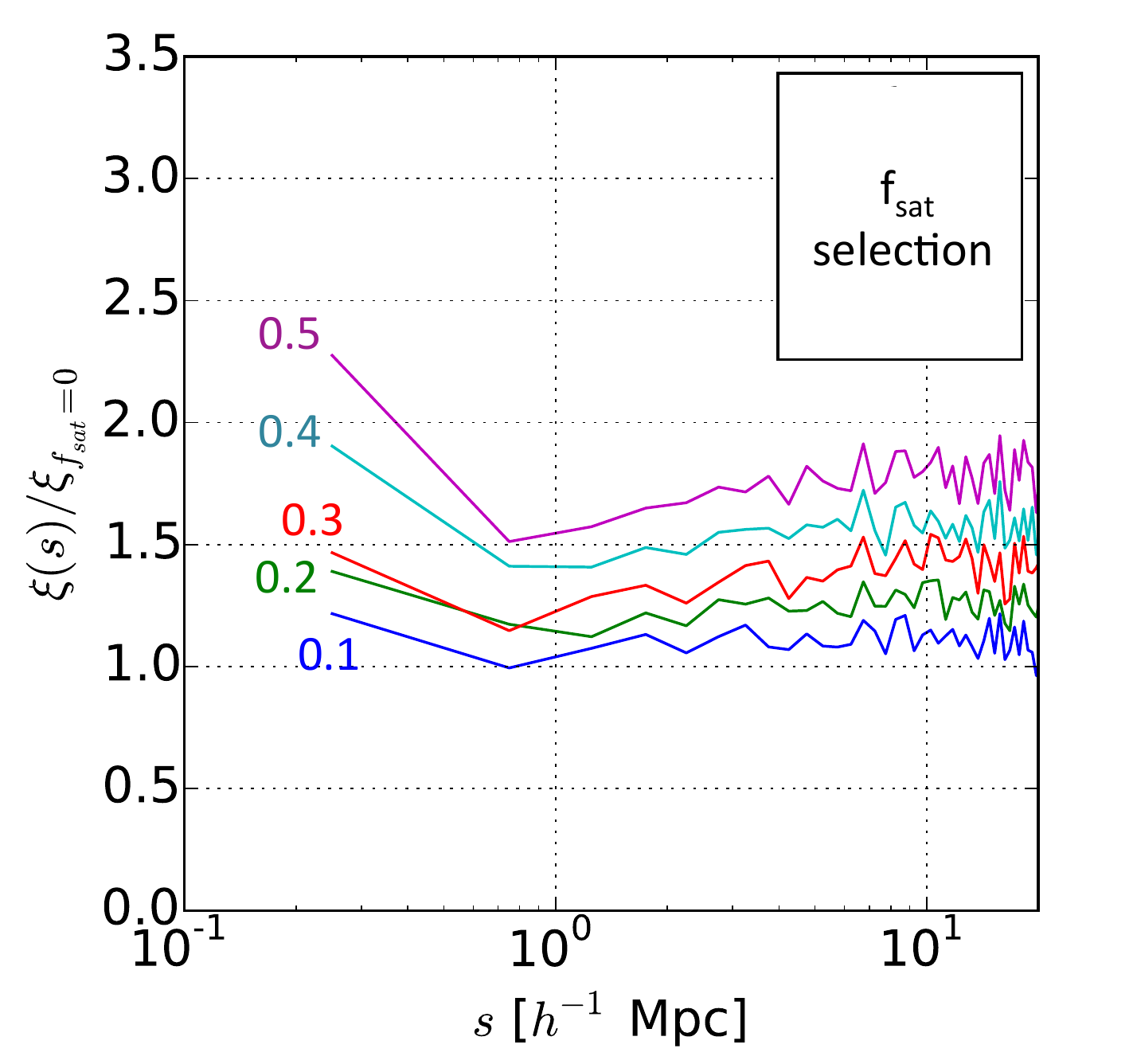}\hfill
\caption{\textit{Left column, top panel}: ratio of the angular correlation functions of the MDPL-LC halos selected by mass, to $w(\theta)$ computed at $M_{mean}=10^{12}h^{-1}$M$_{\odot}$. The curves in the plot go from lower mass (bottom line) to higher mass (top line). \textit{Left column, bottom panel}: ratio of the angular correlation functions of the MDPL-LC halos with varying satellite fraction, to $w(\theta)$ computed at $f_{sat}=0$. The lines in the plot go from lower $f_{sat}$ (bottom line) to higher $f_{sat}$ (top line). \textit{Right column:} same results for the monopole. The top row presents our first experiment (see the text for details) on the lightcone: we impose different halo mass thresholds to the MDPL-LC and apply a standard SHAM. The bottom row displays SHAM in the mass bin $M_h=(1\pm0.5 \times 10^{12}\,h^{-1}$M$_{\odot})$ with varying satellite fractions.}
\label{clustering:plot:para}
\end{center}
\end{figure*}

We parametrize the probability of selecting a halo hosting an ELG as follows:
\begin{equation}
 \begin{aligned}
 P(M_h,M_{mean},\sigma_M,f_{sat})=\hspace{3cm}\\
f_{sat} \mathcal{N}(M_h,M_{mean},\sigma_M,{\rm flag=sat}) +\hspace{1cm}\\
 (1-f_{sat})\mathcal{N}(M_h,M_{mean},\sigma_M,{\rm flag=cen})
\label{eq:probab}
 \end{aligned}
\end{equation}
where $\mathcal{N}$ is a Gaussian distribution with the variable being $M_h$, the halo mass. The parameters are: $M_{mean}$, the mean halo mass of the sample including both host and satellite halos; $\sigma_M$, the dispersion around the mean halo mass; $f_{sat}$, the satellite fraction. The additional parameter ``flag'' enables to identify among the halos the ones that are centrals (flag=cen) or the ones that are satellites (flag=sat).

To qualitatively understand the dependence of clustering on $M_{mean}$ and $f_{sat}$, we impose (i) a maximum halo mass threshold to the MDPL-LC by removing all halos with $M_h>M_{max}$ and we apply the standard SHAM procedure. The higher-mass ($M_{max}>10^{13}\,h^{-1}$M$_{\odot}$) models reproduce well the observed $w(\theta)$, and that the lower-mass models ($M_{max}<10^{13}\,h^{-1}$M$_{\odot}$) match the large-scale clustering, but not the small-scale amplitude witnessed below $\theta\sim0.01^\circ$. The top row in Figure \ref{clustering:plot:para} displays the ratio between the angular (left panel) and the monopole (right) correlation functions of the lower-mass models and the model with $M_{mean}=10^{12}h^{-1}$M$_{\odot}$. We see a mild variation in $w(\theta)$ as a function of the physical scale, and a flatter trend in the monopole.

\begin{figure}
\begin{center}
\hspace{1cm}
\includegraphics[width=8cm]{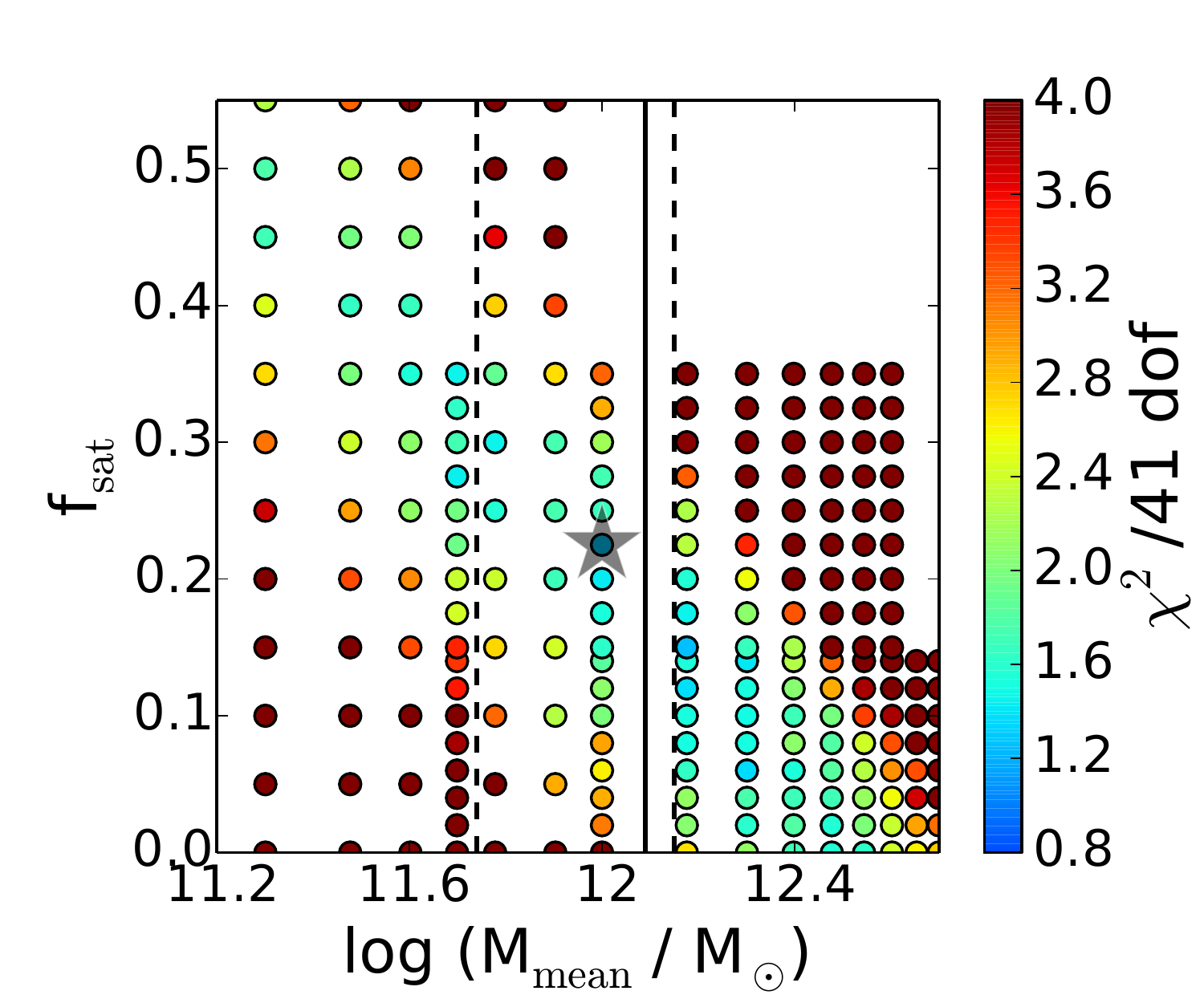}\hfill
\caption{The two parameters driving the model: fraction of satellite ($f_{sat}$) and mean halo mass ($M_{mean}$). The spread around the mean halo mass is fixed at the value $\sigma_M=M_{mean}/2$. The vertical black lines represent the constraints by weak lensing (dashed: lower and upper limits; solid: mean), which rule out the majority of the low-mass and high-mass models. Our best-fit model is highlighted by the star symbol.}
\label{clustering:plot:best:model}
\end{center}
\end{figure}

We next (ii) fix the halo mass by selecting all the halos in the mass bin $M_h=(1\pm0.5\times10^{12}\,h^{-1}$M$_{\odot})$, and vary the satellite fraction. We split this halo catalog into two catalogs, one containing only central halos ($f_{sat}=0$) and one with satellites; then downsample both mocks to match the ELG $n(z)$. 
The bottom panels in Figure \ref{clustering:plot:para} present the variation of the angular and monopole clustering as a function of the scale. At small scales the amplitude of $w(\theta)$ with more than $30\%$ satellite fraction is strongly enhanced compared to the $10-20\%$ cases. In the monopole there is almost no variation with the scale.
We then combine these two products to build galaxy mock catalogs that contain a $f_{sat}$ fraction of satellites (taken from the satellite-only mock) and (1-$f_{sat}$) centrals (from the central-only mock). Satellite fractions between 20\% and 30\% account for the clustering signal on both small and large scales; see Figure \ref{clustering:plot:best:model}.
All the selections above are done on the halo mass defined as $M_{200}$, which correspond to an overdensity threshold of $\Delta_{200}=200\rho_{c}$ \citep{2012MNRAS.423.3018P}, where $\rho_c$ is the critical density of the Universe.

To produce a mock catalog, we randomly select halos from the light-cone according to the probability distribution $P$, defined in Eq. \ref{eq:probab}, until the ELG redshift distribution $n(z)$ in Figure \ref{fig:nz} is achieved. We then construct a grid of mocks by selecting $M_{mean}$ in the range $10^{11.2} - 10^{12.7}\,h^{-1}$M$_{\odot}$, $\sigma_M$ between the values $M_{mean}/ [1., 2., 4. ]$ $h^{-1}$M$_{\odot}$ (the sampling  space is three times larger), and the satellite fraction in the interval $0<f_{sat}<0.5$, to obtain predictions for both $\xi^s$ and $w(\theta)$.
Finally, we compare these model predictions with our measurements by computing a combined $\chi^2$ on scales $2<s<22\,h^{-1}$Mpc for the monopole, and $0.002^\circ<\theta<0.55^\circ$ for the angular clustering, as follows:
\begin{equation}
\chi^2 = \frac{N_\xi \chi_{\xi}^2 + N_w \chi_{w(\theta)}^2}{N_\xi+ N_w},
\end{equation}
where
\begin{equation}
\chi_{w(\theta)}^2 = \frac{1}{N_w}\sum^{N_w}_i \frac{ |w_{observed}(\theta_i)-w_{halos}(\theta_i)|^2 }{\sigma^2 (w_{observed}(\theta_i))},
\end{equation}
and
\begin{equation}
\chi_{\xi}^2 = \frac{1}{N_\xi}\sum^{N_\xi}_i \frac{ |\xi_{observed}(s_i)-\xi_{halos}(s_i)|^2 }{\sigma^2 (\xi_{observed}(s_i))}.
\end{equation}

The possible models accounting for the ELG clustering are degenerate with respect to the mean halo mass and the satellite fraction. In fact, Figure \ref{clustering:plot:best:model} shows that a plethora of $(\log M_{mean}, f_{sat})$ models fit the data: from $(11.3,0.45)$ by $(12,0.2)$ to $(12.5,0)$. Given the 41 degrees of freedom we have, we consider acceptable those models with $\chi^2<1.25$. Models with a higher $\chi^2$ value are rejected at the 90\% level. 

The combination with the weak lensing results breaks this degeneracy and rules out the higher- and lower-mass models. However, among these latter, there is one with $\chi^2=1$ and parameters: $\log M_{mean}=12$, $\sigma_M= M_{mean}/2$, $f_{sat}=22.5\%$ (star symbol in Figure \ref{clustering:plot:best:model}). The angular and redshift-space correlation functions of this best-fit mock are displayed in Figure \ref{clustering:measurement:plot} (blue line), together with the ELG measurements. The weak lensing measurement are perfectly compatible with this best-fit model.

We provide our best-fit MDPL mock catalog to the ELG clustering measurements at \url{http://projects.ift.uam-csic.es/skies-universes/}.


\section{Results and Discussion}
\label{sec:clusteringTrends}

\subsection{ELG clustering trends as a function of magnitude, flux, luminosity and stellar mass}
\label{sec:trends}

\begin{figure*}
\begin{center}
\includegraphics[width=0.33\textwidth]{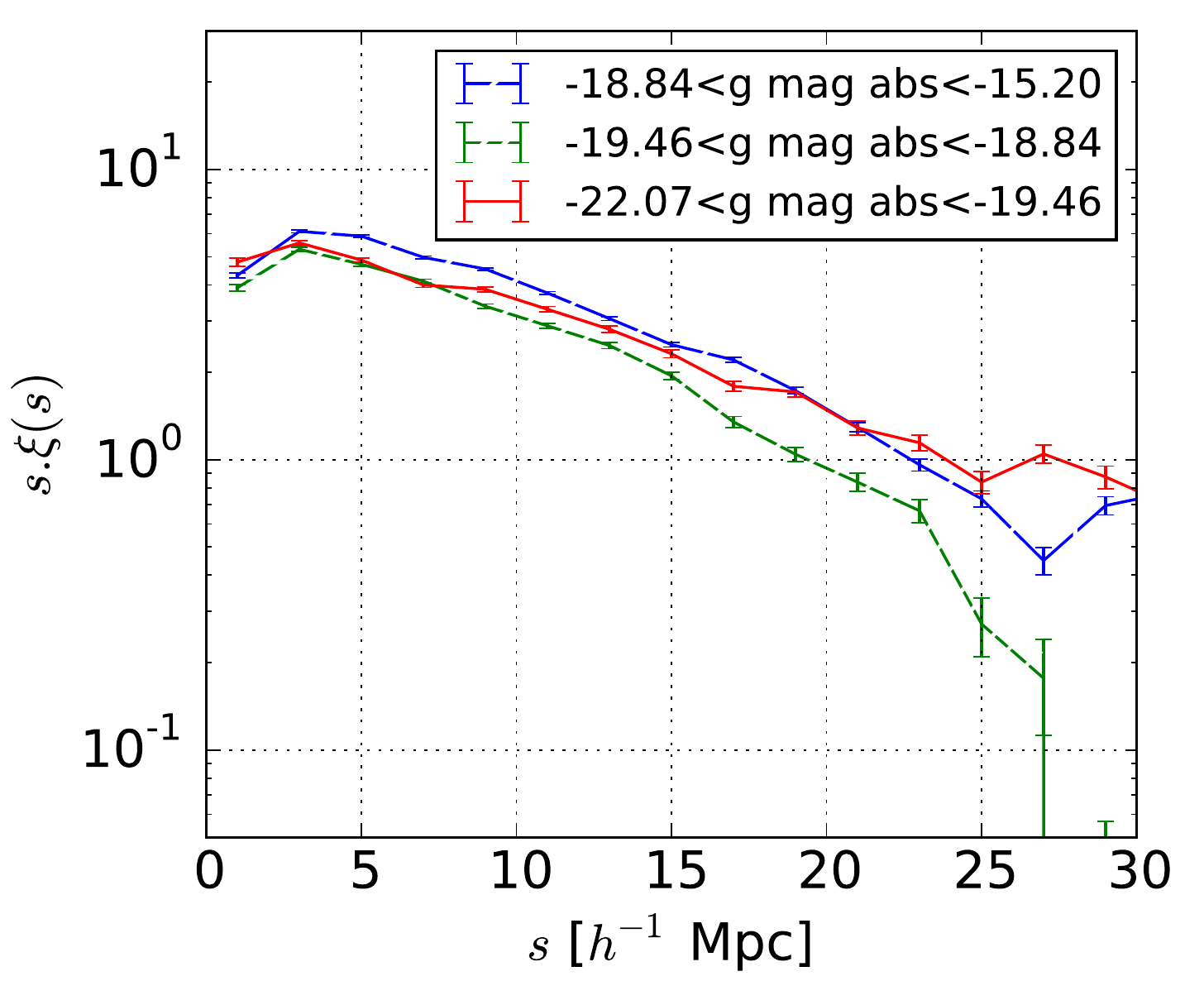}
\includegraphics[width=0.33\textwidth]{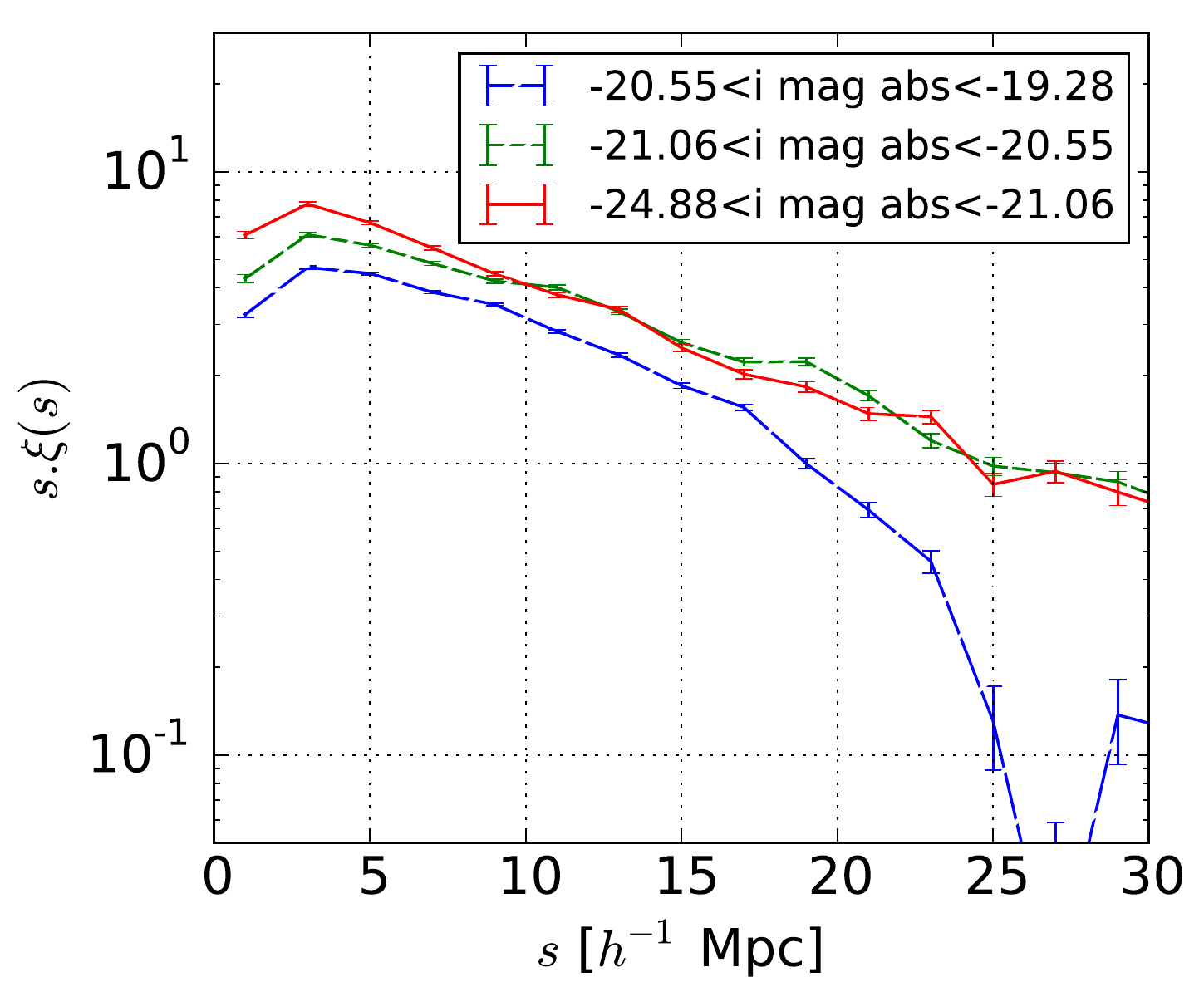}\hfill
\includegraphics[width=0.33\textwidth]{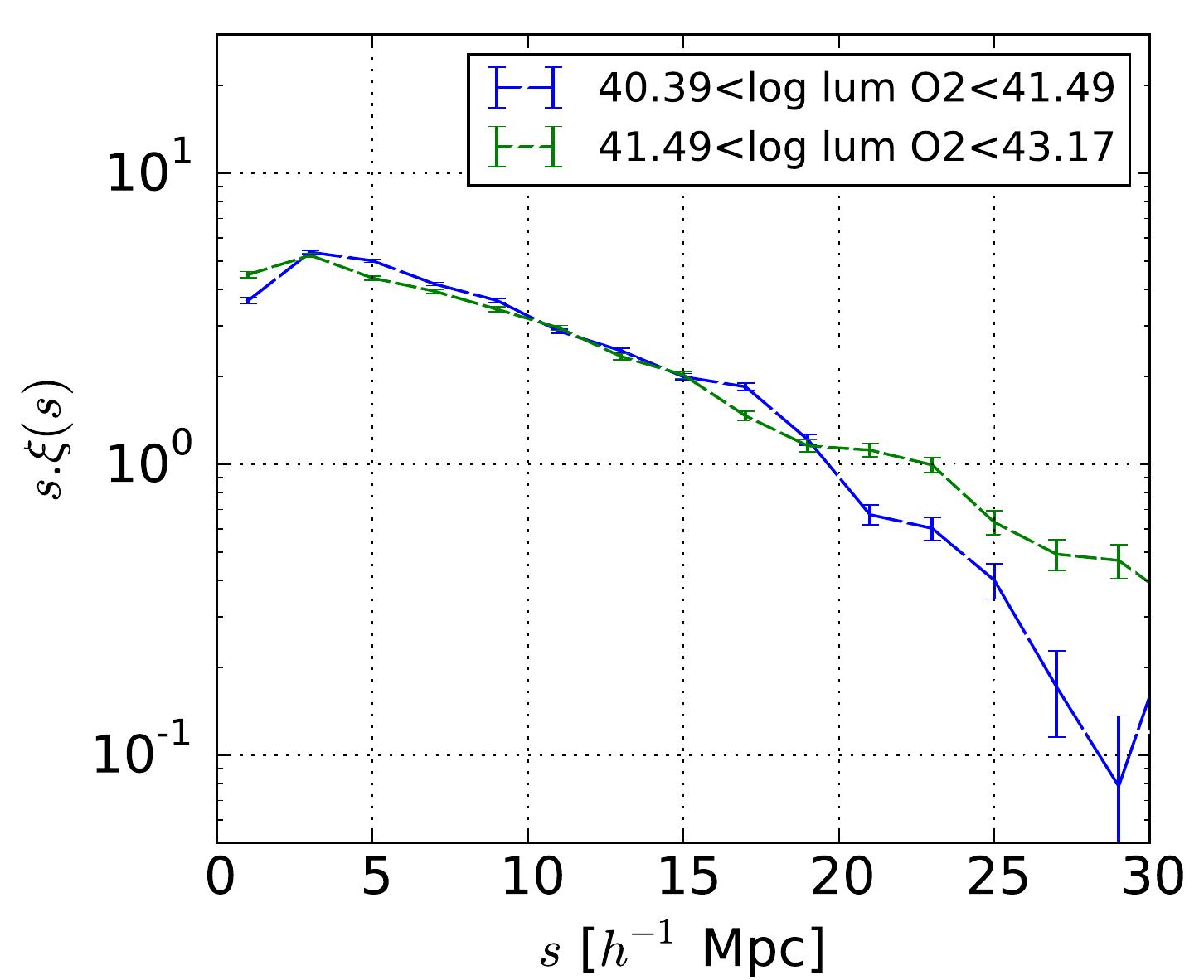} \\
\includegraphics[width=0.33\textwidth]{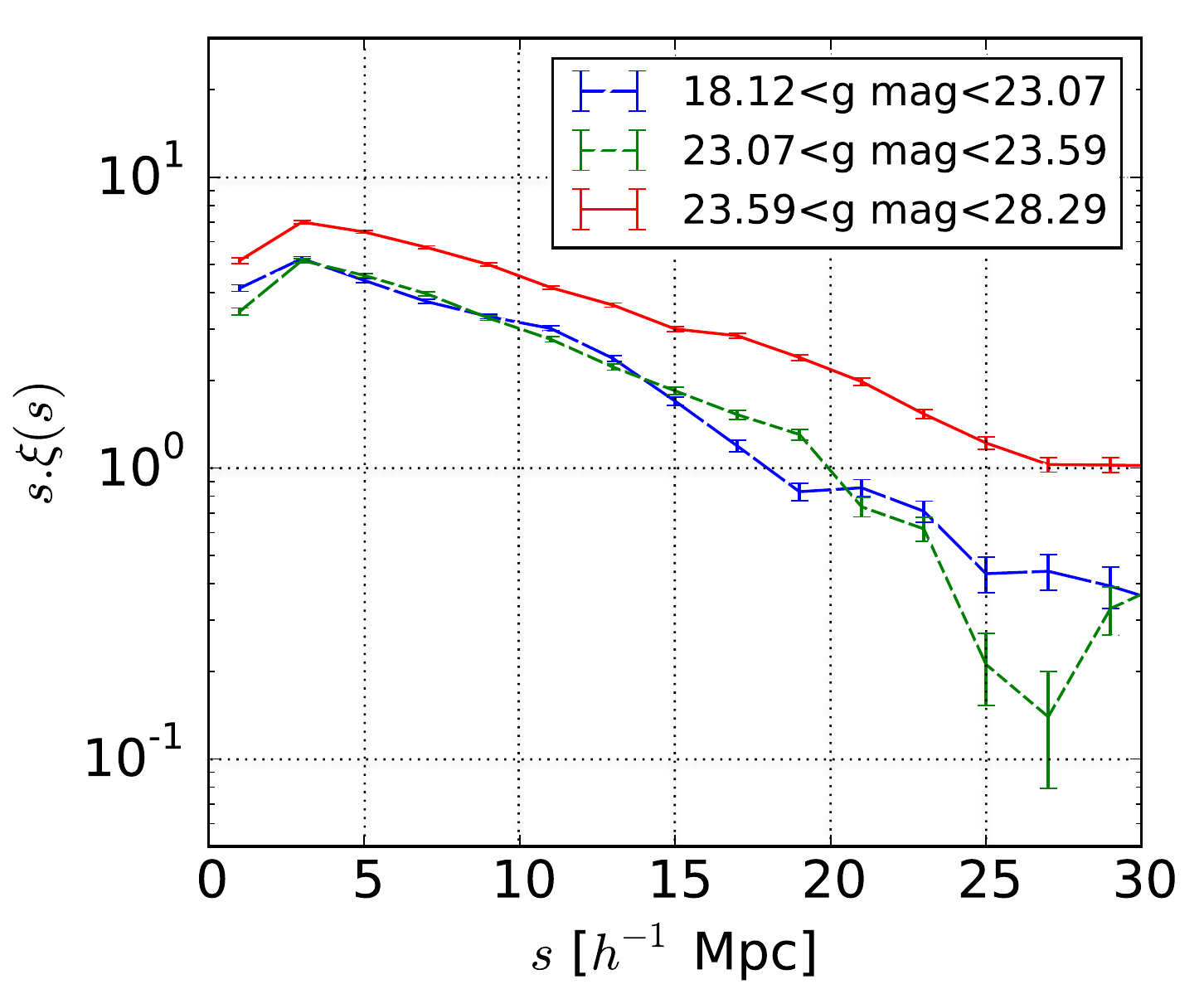}
\includegraphics[width=0.33\textwidth]{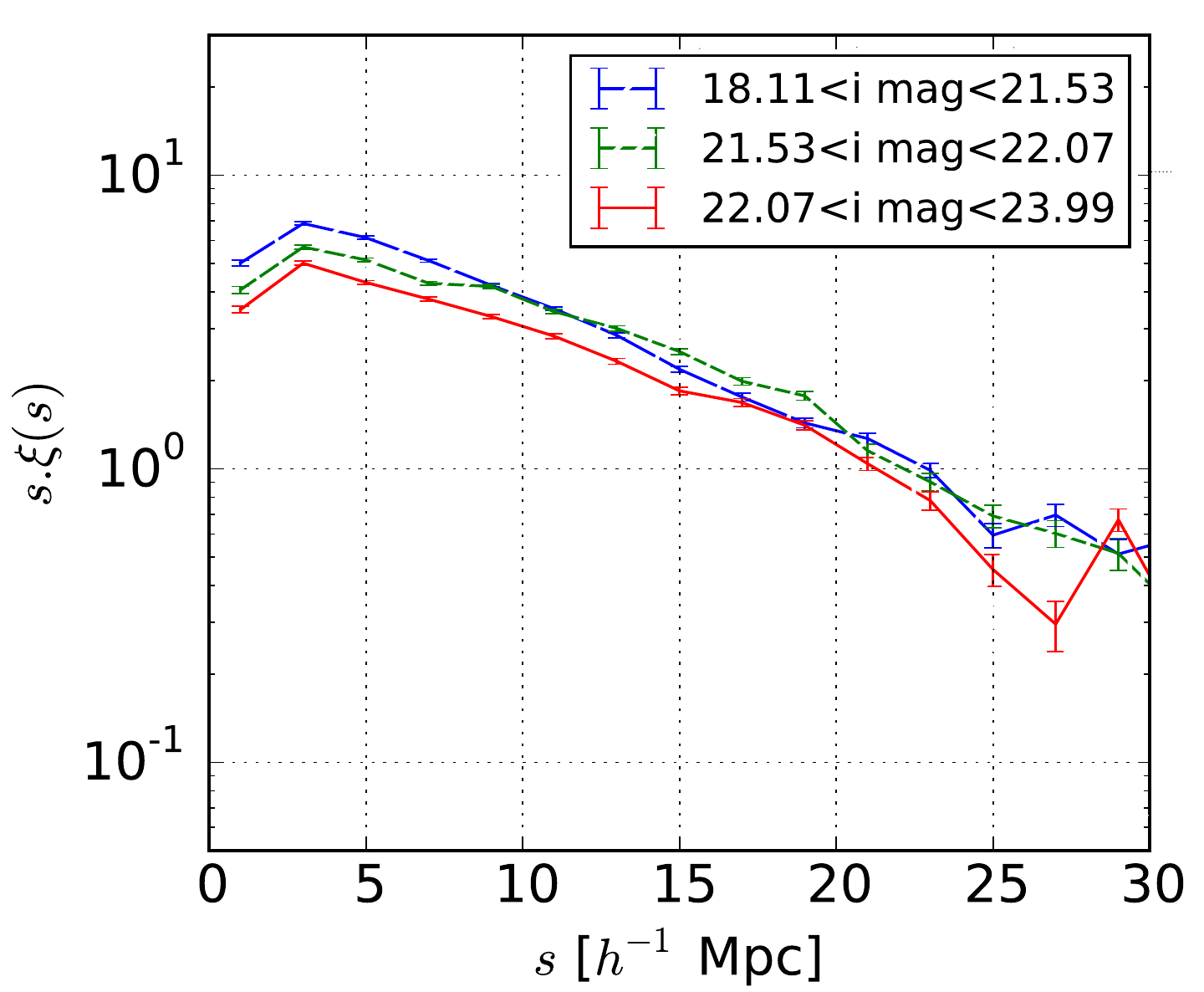}\hfill
\includegraphics[width=0.33\textwidth]{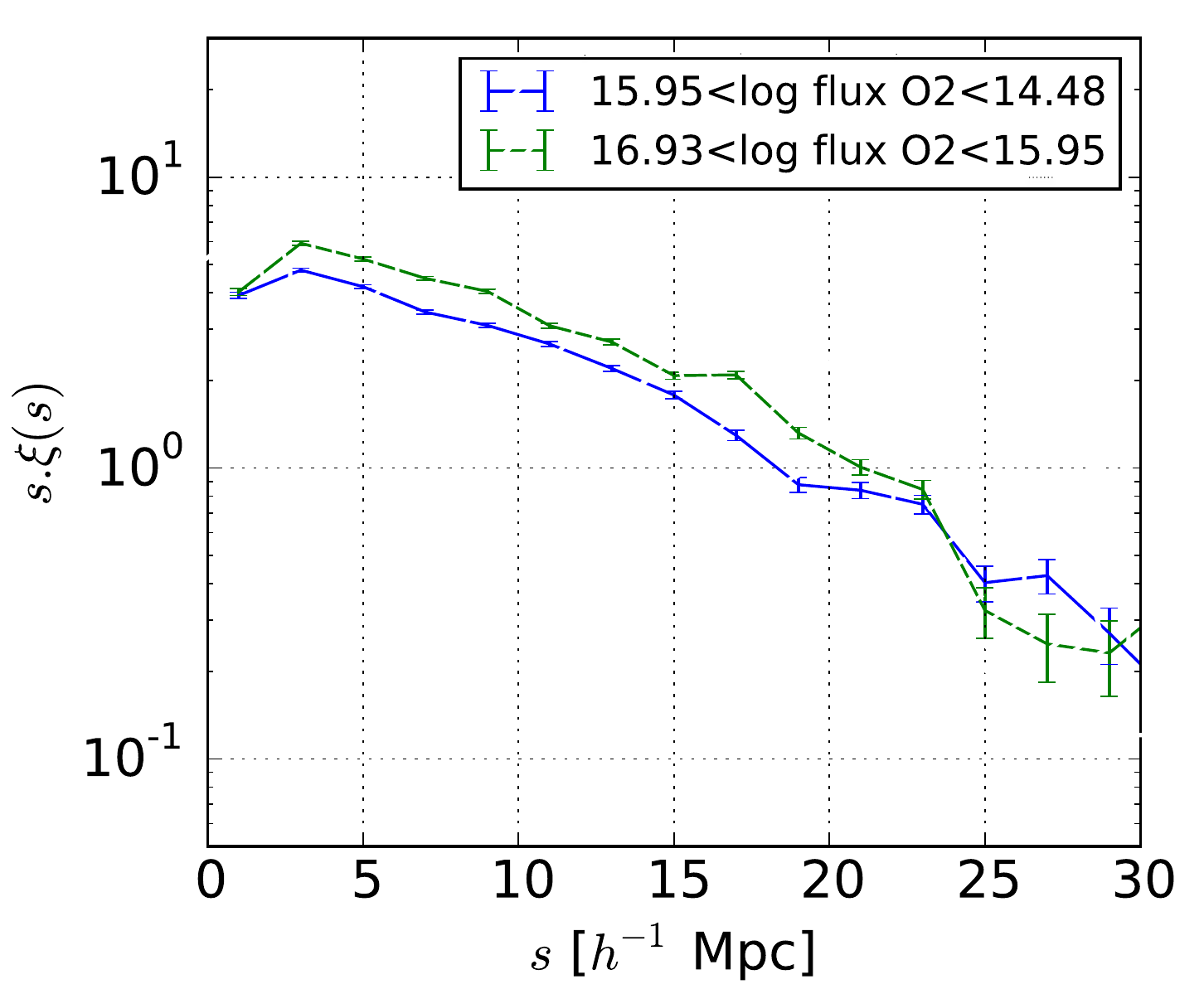}
\caption{VIPERS clustering trends as a function of the $g-$band and $i-$band magnitudes (top row: rest frame; bottom row: observed frame), \OII luminosity (top row) and \OII flux (bottom row). }
\label{fig:xi:trends}
\end{center}
\end{figure*}

We employ the complete VIPERS data sample at $z\sim0.8$, which has about $30,000$ reliable redshifts in the range $0.6<z<1$, to investigate trends of the clustering amplitude (bias) with observed or rest frame broad band magnitude or emission line flux. 
To this purpose, we measure the emission line properties in the VIPERS spectra and find a significant \OII flux in about two thirds of them; the rest does not show emission lines (Comparat et al., in prep.). We bin the data according to apparent and absolute magnitude, \OII flux and luminosy, and measure the clustering in each sample (the binning scheme was set to contain between $9000$ and $10,000$ data points). Figure \ref{fig:xi:trends} shows our ELG results in the observed (bottom row) and rest frame (top row).
Consistently with previous analyses \citep[e.g.,][]{2013A&A...557A..17M, 2013ApJ...767...89M}, we find that the brighter the selection in the $i$-band, either observed or rest-frame, the higher the bias. Analogously, the fainter the $g$-band limit, either observed or rest-frame, the higher the bias. The anti-correlation between \OII flux and bias is only seen in the observed frame (the difference is $\sim1.4$); in the rest frame it is not significant. It would be interesting to further investigate the correlation between \OII luminosity and $g$-band magnitude in the small-scale clustering, but with the resolution of current data we are not able to push the analysis to scales $\sim 200h^{-1}$kpc, which is the typical virial radius of a halo of mass $10^{12}h^{-1}$M$_{\odot}$. New data from eBOSS will be able to address this issue.
The results above indicate that if we have a $g$-selected ELG sample and \OII fluxes for a certain number of its galaxies, in order to maximize its clustering signal, we should select the ELGs with brighter $i$-band magnitudes. 
To investigate the clustering dependence on stellar mass, we map the host halo masses for ELGs at $z\sim0.8$, $M_h\sim10^{12}h^{-1}$M$_{\odot}$, onto stellar mass values using the stellar-to-halo-mass relation by \cite{2012ApJ...744..159L}, see their Figure 11. Our data are right before the ``knee'' at $M_{\star}\sim3.5\times 10^{10}h^{-1}$M$_{\odot}$.


\subsection{Star formation efficiency}
\label{sec:SF_efficiency}

\begin{figure*}
\begin{center}
\includegraphics[width=0.8\textwidth]{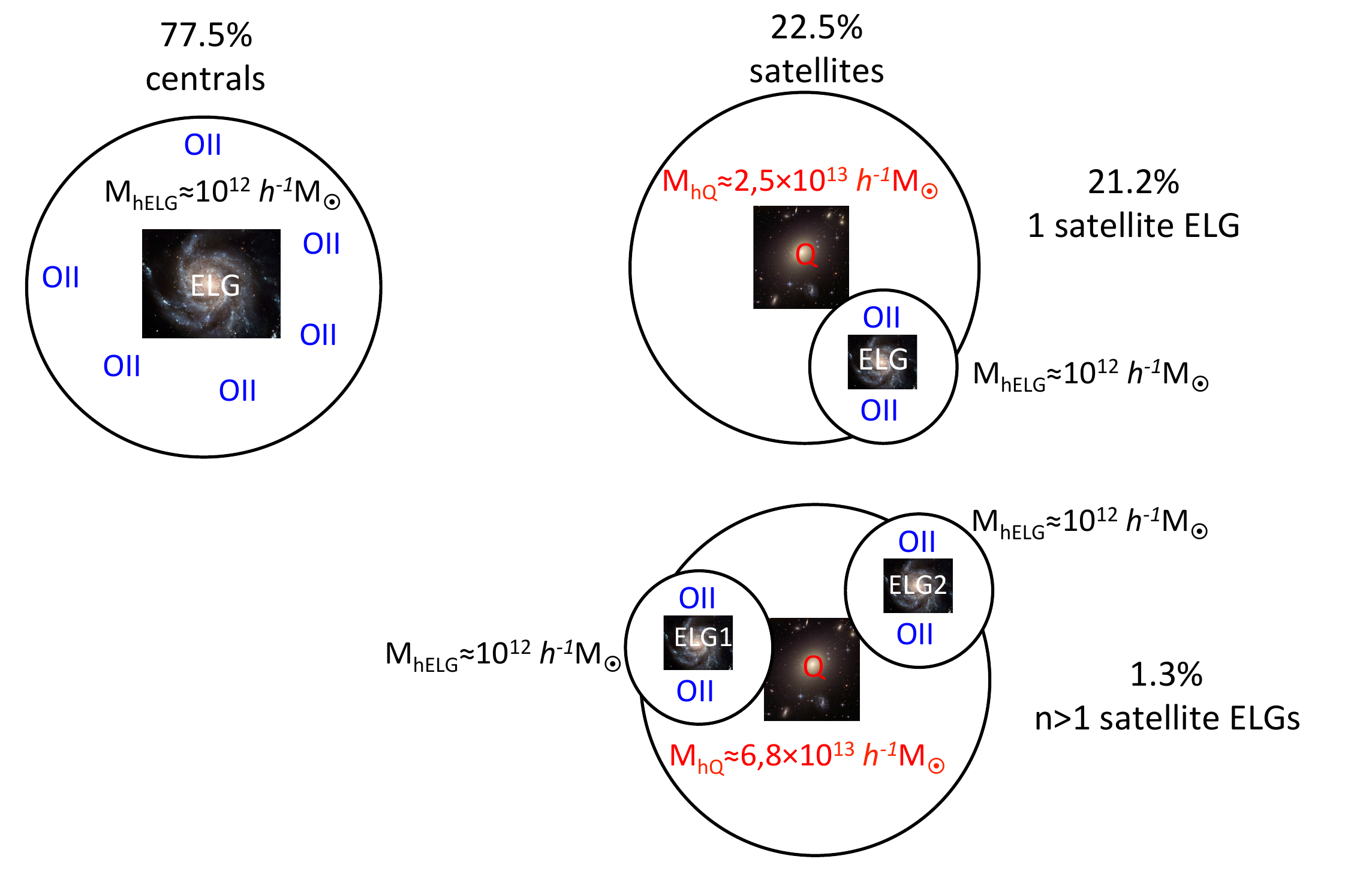}
\caption{Schematic diagram of possible ELG configurations. ELGs at $z\sim0.8$ typically live in halos of mass $M_{h}\sim(1\pm 0.5)\times10^{12}\,h^{-1}$M$_{\odot}$ and $22.5\%$ are satellites belonging to larger halos, whose central galaxy is quiescent. Among these satellite configurations, $21.2\%$ of parent halos with $M_{hQ}\sim2.5\times 10^{13}\,h^{-1}$M$_{\odot}$ host one satellite ELG, and only $1.3\%$ of parents host more than one satellite ELG. The maxium number of satellites, $n=1.8$, is achieved in the highest-mass case, $M_{hQ}\sim6.8\times 10^{13}\,h^{-1}$M$_{\odot}$. See the text for details.}
\label{fig:cartoon}
\end{center}
\end{figure*}

From our analysis, the typical halo masses hosting ELGs at $z\sim0.8$ are $M_h\sim(1\pm 0.5)\times10^{12}\,h^{-1}$M$_{\odot}$, and 22.5\%$\pm2.5$\% of them are satellites belonging to a larger halo, whose central is a quiescent galaxy. Figure \ref{fig:cartoon} provides a schematic representation of the possible ELG configurations. A total of $22.1\%$ ELGs are single satellites belonging to a parent halo with mass $M_{hQ}\sim2.5\times 10^{13}\,h^{-1}$M$_{\odot}$; only in $1.3\%$ of the cases the parent halo hosts more than one satellite ELG. The maximum number of satellites, $n=1.8$, is achieved in the highest-mass case, where $M_{hQ}\sim6.8\times 10^{13}\,h^{-1}$M$_{\odot}$. These results imply that the mean number of ELG satellites is only slighlty larger than unity ($\sim1.01$). The quiescent galaxies at the center of the parent halos are not included in the sample, since the stellar masses for ELGs from the SHMR discussed above are too low for halos of $10^{13}\,h^{-1}$M$_{\odot}$. 

The typical masses for halos hosting ELGs suggest that we are sampling halos ($\sim10^{12}\,h^{-1}$M$_{\odot}$) that form stars in the most efficient way, according to the stellar-to-halo mass ratio discussed by \cite{2013ApJ...762L..31B} (see their Figure 1, bottom panel). This result opens a new science field and, hopefully, in the near future, integrated models combining N-body simulations with semi-analytic models (SAMs) will be able to probe star formation and shed some light on the correlations between \OII flux and magnitude in the clustering of galaxies.


\section{Summary}
\label{sec:summary}

We have presented an analysis of the halo occupation distribution for emission line galaxies, which jointly accounts for three measurements: the angular correlation function, the monopole, and the weak lensing signal around ELGs (see Section \ref{sec:clustering}). Our procedure can be summarized in the following points:
\begin{itemize}
\item Apply the SUGAR (Rodriguez-Torres et al. (2015), in prep.) algorithm to the 11 snapshots available from the MDPL simulation to construct a light-cone (Section \ref{sec:MD_simulation}), with the same geometry and angular footprint of the ELG data. \\

\item Modify the traditional SHAM technique (Section \ref{sec:SHAM_procedure}), to account for the ELG incompleteness, by selecting model galaxies by mass, until we match the observed ELG $n(z)$. In this way, our mock is constrained by the observed ELG redshift distribution, and represents a reliable model.\\

\item Parametrize the probability of selecting a halo hosting a ELG with Eq. \ref{eq:probab}, in terms of the mean halo mass of the sample ($M_{mean}$), the dispersion around the mean ($\sigma_{M}$), and the satellite fraction ($f_{sat}$). The additional parameter ``flag'' enables to distinguish central and satellite halos.\\

\item We perform two experiments (see Section \ref{sec:SHAM_procedure}) on the MDPL light-cone to derive information on which are the halo mass and satellite fraction ranges of values we need to input in our modified SHAM model to correctly fit the ELG clustering signal. \\

\item Construct a grid of models based on these values, and jointly fit both angular and redshift-space clustering (see Section \ref{sec:SHAM_procedure}). 
Our best-fit models (see Figure \ref{clustering:plot:best:model}) are degenerate with respect to $M_{mean}$ and $f_{sat}$. The combination with the weak lensing analysis (see Section 3.1) breaks this degeneracy and rules out the highest and lowest mass models. 
Our best-fit ($\chi^2=1$) model is shown in Figure \ref{clustering:measurement:plot} together with the ELG measurements, and is given by $\log M_{mean}=12$, $f_{sat}=22.5\%$, $\sigma_M=M_{mean}/2$.
\end{itemize}

To conclude, we have built and released to the community a reliable galaxy mock catalog that correctly fits the clustering amplitude of the $ugri$ ELG sample constructed by matching spectroscopic redshifts from BOSS DR12, VIPERS and DEEP2 (for details see Section \ref{sec:dataall}). With these tools, we can begin building many realizations of the density field to predict errors on the BAO measurement. 

The measured halo masses for halos hosting emission-line galaxies indicate that we are sampling the halos that form stars in the most efficient way, according the stellar-to-halo mass ratio discussed by \cite{2013ApJ...762L..31B} (see their Figure 1, bottom panel). This is an important point for the future, and opens the path to further studies to understand the correlation between clustering and the strength of emission lines. With the resolution available from current data, we are not able to push the analysis to the typical scales ($\sim200h^{-1}$kpc) of halos of $10^{12}\,h^{-1}$M$_{\odot}$; however, next-generation surveys, as eBOSS and DESI, will provide better resolution, and in the near future we should be able to build robust combinations of N-body simulations and SAMs that will address those questions.
 

\section*{Acknowledgments}

GF is supported by the Ministerio de Educaci\'{o}n y Ciencia of the Spanish Government through FPI grant AYA2010-2131-C02-01.
JC acknowledges financial support from MINECO (Spain) under project number  AYA2012 - 31101. 
GY acknowledges financial support from MINECO (Spain) under project number  AYA2012-31101 and grant FPA2012-34694. 
EJ acknowledges the support of CNRS, and the Labex OCEVU.
SEN acknowledges support by the Deutsche Forschungsgemeinschaft under the grant NU 332/2-1.
GF, FP, SART, AK, SN and CC acknowledge financial support from the Spanish MICINN Consolider-Ingenio 2010 Programme under grant MultiDark 
CSD2009 - 00064, MINECO Centro de Excelencia Severo Ochoa Programme under grant SEV-2012-0249, and MINECO grant AYA2014-60641-C2-1-P.
GF, JC and FP wish to thank the Lawrence Berkeley National Laboratory for the hospitality during the creation of this work.
FP acknowledges the spanish MEC ``Salvador de Madariaga'' program, Ref. PRX14/00444.

The MultiDark Planck  simulation  has  been performed in the Supermuc supercomputer at
the Libniz Supercomputing Center (LRZ, Munich) thanks to the cpu time awarded by PRACE (proposal number 2012060963).

Funding for SDSS-III has been provided by the Alfred P. Sloan Foundation, the Participating Institutions, the National Science Foundation, and the U.S. Department of Energy Office of Science. The SDSS-III web site is http://www.sdss3.org/.

SDSS-III is managed by the Astrophysical Research Consortium for the Participating Institutions of the SDSS-III Collaboration including the University of Arizona, the Brazilian Participation Group, Brookhaven National Laboratory, Carnegie Mellon University, University of Florida, the French Participation Group, the German Participation Group, Harvard University, the Instituto de Astrofisica de Canarias, the Michigan State/Notre Dame/JINA Participation Group, Johns Hopkins University, Lawrence Berkeley National Laboratory, Max Planck Institute for Astrophysics, Max Planck Institute for Extraterrestrial Physics, New Mexico State University, New York University, Ohio State University, Pennsylvania State University, University of Portsmouth, Princeton University, the Spanish Participation Group, University of Tokyo, University of Utah, Vanderbilt University, University of Virginia, University of Washington, and Yale University. 

\bibliographystyle{mn2e}
\bibliography{./bibliography}

\end{document}